# SCOTCH – Search for Clandestine Optically Thick Compact HIIs ★

A. L. Patel,[1]† J. S. Urquhart,[1]‡ A. Y. Yang,[3,4] T. J. T Moore,[5] K. M. Menten,[2] M. A. Thompson,[6] M. G. Hoare,[6] T. Irabor,[6] S. L. Breen,[7] M. D. Smith[1]

[1] *Centre for Astrophysics and Planetary Science, University of Kent, Canterbury, CT2 7NH, UK*
[2] *Max Planck Institute for Radio Astronomy, Auf dem Hügel 69, 53121 Bonn, Germany*
[3] *National Astronomical Observatories, Chinese Academy of Sciences, A20 Datun Road, Chaoyang District, Beijing 100101, P. R. China*
[4] *Key Laboratory of Radio Astronomy and Technology, Chinese Academy of Sciences, A20 Datun Road, Chaoyang District, Beijing, 100101, P.R. China*
[5] *Astrophysics Research Institute, Liverpool John Moores University, Twelve Quays House, Egerton Wharf, Birkenhead CH41 1LD, UK*
[6] *School of Physics and Astronomy, University of Leeds, Leeds, UK*
[7] *Sydney Institute for Astronomy (SIfA), School of Physics, University of Sydney, NSW 2006, Australia*

Last updated 2022 June 10; in original form XXX XXXX XX

**ABSTRACT**
This study uses archival high frequency continuum data to expand the search for Hypercompact HII regions and determine the conditions at which they appear, as this stage high mass star formation is short-lived and rare. We use 23 GHz continuum data taken towards methanol masers, which are an excellent signpost for very young embedded high-mass protostars. We have searched for high-frequency, optically thick radio sources to identify HC HII region candidates. The data cover 128 fields that include 141 methanol masers identified by the Methanol Multibeam (MMB) survey. We have detected 68 high-frequency radio sources and conducted a multi-wavelength analysis to determine their nature. This has identified 49 HII regions, 47 of which are embedded in dense clumps fourteen of which do not have a 5 GHz radio counterpart. We have identified 13 methanol maser sites that are coincident with radio sources that have a steep positive spectral index. The majority of these are not detected in the mid-infrared and have been classified as protostellar or young stellar objects in the literature and we therefore consider to be good HC HII region candidates, however, further work and higher resolution data are needed to confirm these candidates.

## 1 INTRODUCTION

Massive stars (>8 $M_\odot$) play a vital role in regulating the efficiency of star formation in clusters, due to their powerful outflows, strong stellar winds and chemical enrichment of materials in the interstellar medium (ISM; McKee & Ostriker 2007). Despite their importance, we do not currently have a detailed understanding of the processes that enable massive stars to form. This is primarily due to the fact that these objects are rare and evolve rapidly (their Kelvin-Helmholtz time-scale is much shorter than the free-fall time of the host clump). A consequence of this is that they reach the main sequence while still deeply embedded in their natal molecular cloud. The earliest stages in their evolution, therefore, take place behind hundreds of magnitudes of visual extinction and can only be studied at far-infrared (FIR) and submillimetre wavelengths (see the review paper by Churchwell 2002 and references therein).

There have been many reviews proposing an evolutionary scheme for the formation of high mass stars (i.e. Molinari et al. 1996; Evans 1999; McKee & Ostriker 2007; Zinnecker & Yorke 2007; Motte et al. 2018). The basic evolutionary scheme that has emerged is as follows: Embedded within clumpy molecular clouds are dense starless clumps. These begin to collapse and fragment resulting in the formation of protostars. These protostellar objects are surrounded by an accretion disc that is associated with a bipolar outflow. Outflows are driven by powerful jets that inject momentum and entrain large quantities of the surrounding molecular material (Purser et al. 2016). When a protostellar object becomes massive enough, it transitions into a massive young stellar object (MYSO), which is the next evolutionary phase of high mass star formation. At this point newly formed MYSOs can be detected and classified based on their near-infrared (NIR) and mid-infrared (MIR) spectral energy distributions (SEDs; Lumsden et al. see 2013, for more details).

As the MYSOs continue to evolve, the circumstellar environ-

★ The full version of Tables 3 and 4 are only available in electronic form at the CDS via anonymous ftp to cdsarc.u-strasbg.fr (130.79.125.5) or via http://cdsweb.u-strasbg.fr/cgi-bin/qcat?J/MNRAS/.
† E-mail: alp48@kent.ac.uk
‡ E-mail: j.s.urquhart@kent.ac.uk

2    *A.L Patel et al.*

ment begins to change. Their mass and luminosity increase, which leads to a rise in surface temperature, which produces copious amounts of Lyman continuum photons. These high energy photons ionize the surrounding material, resulting in the formation of bubbles of hot ionized gas, referred to as H II regions. These regions are over-pressured with respect to the surrounding gas and are therefore expected to expand rapidly, driving ionisation shocks into the ambient medium and compressing the surrounding natal material (Dyson et al. 1995) and potentially initiating the formation of a new generation of stars (e.g., Whitworth et al. 1994; Thompson et al. 2012).

Initially, ultracompact (UC) H II regions were thought to represent an intermediate stage between a MYSO and a more evolved optically visible H II region (Wood & Churchwell 1989), however, a smaller, more compact and optically thick type of H II region was discovered by Gaume et al. (1995). Their size and density indicate that these are even younger examples of H II regions that are distinct from the more evolved UC H II regions; these are known as hypercompact (HC) H II regions. The current observationally derived stages in the H II region sequence consists of three stages, HC H II region ($n_e > 10^6$ cm$^{-3}$ and diameter < 0.03 pc; Kurtz & Hofner 2005), followed by, UC H II region ($n_e > 10^4$ cm$^{-3}$ and diameter < 0.1 pc; Wood & Churchwell 1989), compact H II region ($n_e > 10^4$ cm$^{-3}$ and diameter > 0.1 pc; Wood & Churchwell 1989) after which the H II region breaks out of its natal cloud and evolves into a classical H II region, becoming visible at optical wavelengths. It is important to note that these stages are not necessarily distinct stages of evolution but rather represent a continuum of evolution (e.g., see Fig. 9 from Yang et al. 2021) as these objects are subject to an observational bias depending on the resolution and frequency used.

Theoretically, there are many different models predicting the outcomes for the evolution of H II regions. For example, turbulent core and ionisation feedback models such as McKee & Tan (2003) envisage that the youngest H II regions expand into outflow driven cavities, away from the accretion flows, implying the early development of H II regions is part of the star formation process. Whereas the models of Hosokawa et al. (2010) suggest that the high accretion rates of material onto the star causes its outer layers to swell, thus reducing the star's effective temperature. This implies that the development of H II regions does not occur until after the accretion phase has finished (Yang et al. 2019). The processes of the early stages of high-mass star formation, still remain an open question, which is mainly due to the relatively low statistics we have to date.

A feature of low- (Anglada et al. 1996) and high-mass (Purser et al. 2016) star formation are collimated, high-velocity (≥ 300 km s$^{-1}$), ionized jets of materials that are ejected along the rotation axis of accretion discs. From a theoretical standpoint, the link between discs and jets is crucial for the formation of high-mass stars to prevent powerful stellar radiation from sweeping away the natal cocoon before too much mass is gained by the protostar. Radio jets and H II regions can often be mistaken for each other due to their similar spectral indices. From recent observations, most protostellar radio jets produce free-free radio emission and spectral indices that range between $-0.1 \leq \alpha \leq 1.6$, which is similar to that of compact and optically thick H II regions (Purser et al. 2016), however, they are associated with significantly lower Lyman photon fluxes making jets and H II regions easy to distinguish. Jets and outflows are an important part of the star formation process as they carry away excess angular momentum. Additionally, they provide a source of momentum to drive large-scale molecular outflows (Beuther et al. 2002). Detecting and understanding this stage of star formation may aid in differentiating between competitive/Bondi-Hoyle accretion (Bonnell et al. 2001) and turbulent core accretion models (McKee & Tan 2002).

**Table 1.** Summary of the MMB survey papers and the number of 6.7 GHz methanol maser sites detected broken down by Galactic longitude range. The total number of maser sites detected is 972 between $186° \leq l \leq 60°$.

| Galactic Longitude range | No. of maser sites | Reference |
|---|---|---|
| $345° \leq l \leq 6°$ | 183 | Caswell (2010) |
| $6° \leq l \leq 20°$ | 119 | Green et al. (2010) |
| $330° \leq l \leq 345°$ | 198 | Caswell et al. (2011) |
| $186° \leq l \leq 330°$ | 207 | Green et al. (2012) |
| $20° \leq l \leq 60°$ | 265 | Breen et al. (2015) |

Given that HC H II regions provide the earliest indication that the star is beginning to ionize its surroundings, they represent a key stage in the development of massive stars. They manifest themselves as expanding bubbles of hot ionized gas that impede further accretion and set the final mass of the star. They represent an important phase in the early evolutionary stages of massive stars and their environments. However, HC H II regions are optically obscured by a thick cocoon of dust making them difficult to detect for three primary reasons. First, their small sizes results in low flux densities. Second, most radio surveys have been conducted at frequencies between 1 and 8 GHz (e.g., THOR between 1−2 GHz; Beuther et al. 2016; Bihr et al. 2016, CORNISH at 5 GHz; Hoare et al. 2012; Purcell et al. 2013; Irabor et al. 2023) and GLOSTAR (Global view of Star formation) at 4−8 GHz; Brunthaler et al. 2021; Medina et al. 2019; Dzib et al. 2023) that are optimized for objects with electron densities of $\sim < 10^4$ cm$^{-3}$ and so are not sensitive to the higher-density HC H II regions that are optically thick at radio frequencies below 20 GHz (Kurtz & Hofner 2005). Third, the HC H II phase is likely to be brief (< $10^5$ years) and therefore, rare (Comeron & Torra 1996; González-Avilés et al. 2005).

Currently, only 23 HC H II regions have been identified (Yang et al. 2019; Yang et al. 2021 and references therein) and most of these have been serendipitously detected during high-frequency radio observations towards known UC H II regions. The current sample is therefore limited in number and the manner in which its members have been identified means that they may not be fully representative of this stage in the evolution of H II regions. Consequently, our lack of understanding of their properties makes it difficult to constrain theoretical models for the early evolution of massive stars or to determine if HC H II regions are physically distinct from UC H II regions. It is therefore crucial to increase the sample size of HC H II regions and to do this in a way that avoids relying on low-frequency radio surveys to identify candidates, which are likely to result in a bias to very bright and/or more evolved HC H II regions.

6.7 GHz class II methanol masers are known to be excellent tracers of embedded massive stars in the early stages of their evolution (Breen et al. 2013). Class II methanol masers are radiatively pumped by the strong mid-infrared emission emitted by the closest environment of MYSOs (Cragg et al. 1992). A study of the properties of molecular outflows associated with methanol masers conducted by de Villiers et al. (2015) revealed that methanol masers appear after the outflow phase once accretion onto the protostar is strong enough to generate sufficient luminosity to begin warming the envelope (Urquhart et al. 2022; Billington et al. 2019). A recent targeted 5 GHz radio continuum study by Hu et al. (2016) found that ∼30 per cent of methanol masers are associated with compact radio emission, however, a more recent survey study by (Nguyen et al. 2022) reported a lower association rate of 12 per cent based on the unbiased Global View of Star Formation in the Milky Way (GLOSTAR) survey (Brunthaler et al. 2021).

**Table 2.** Antennae configurations and the theoretical sensitivities at 23.7 GHz for the different array configurations used, where 23.7 GHz is the mid-point of the continuum bands. The primary beam at this frequency is ∼2.0 arcmin. The sensitivities are based on an average integration time of 6 minutes per source and an average Declination of $-36°$.

| Array configurations | Synthesised beam (arcsec) | Baseline Min. (m) | Baseline Max. (m) | Largest angular scale (arcsec) | Sensitivity rms (mJy beam$^{-1}$) |
|---|---|---|---|---|---|
| EW367 | 4.83 × 8.22 | 46 | 367 | 35 | 0.41 |
| H214  | 5.52 × 6.74 | 82 | 247 | 50 | 0.35 |
| H168  | 7.26 × 9.21 | 61 | 192 | 60 | 0.35 |
| 750A  | 2.11 × 3.60 | 77 | 735 | 20 | 0.42 |

In a recent study, Jones et al. (2020) characterised MYSOs hosting 6.7 GHz methanol masers from the MMB and find almost 73 per cent of class II methanol masers are associated with a compact source in all four Hi-Gal far-infrared and submillimeter bands (70-500 $\mu$m; Molinari et al. 2010b). The flux at these wavelengths results from re-emission of young MYSOs' strong UV radiation by the warm surrounding dust, whose mid-IR radiation pumps the class II methanol masers. These masers are therefore a useful way to identify a population of young high-mass protostars prior to the formation of a UC H II region, although some class II methanol masers are found in the envelopes of UC H II regions, e.g., the archetypical UC H II W3(OH), where they coexist with OH masers (Menten et al. 1992)

Combined these studies provide evidence that the methanol maser phase starts shortly after the formation of a high-mass protostar and persists until after the formation of a HC H II region, but is disrupted as the H II region expands into an UC H II and compact H II region (cf. Walsh et al. 1998). Methanol masers are likely to be associated with the early stages of high-mass protostellar objects (>3 $M_\odot$; Minier et al. 2003), prior to the formation of an optically thin H II region and are therefore an excellent starting point to search for new HC H II regions.

In this work we present the unexploited 23.7 GHz continuum data obtained by Titmarsh et al. (2014, 2016). Their observations were carried out to investigate the relationship between water masers and 6.7 GHz class II methanol masers in the Methanol MultiBeam (MMB) survey overview; (Green et al. 2009; see Table 1 for details). We use these high frequency data to search for HC H II region candidates associated with the high-mass protostellar objects signposted by the methanol masers. This paper is organised as follows: Section 2 describes the observational setup and reduction procedures used to process and image our data. In Section 3 we present the detection statistics and use multi-wavelength data to determine the nature of these radio sources and identify H II regions. In Section 4 we derive physical properties for the H II regions and identify those associated optically thick nebulae and distinguish between those likely to be HC H II regions and radio jets. In Section 5 we discuss the evidence supporting these classifications. Finally, in Section 6 we summarize our results and highlight the main findings.

## 2 OBSERVATIONS AND DATA REDUCTION

### 2.1 Archival data

The observations used in the study were made with the Australia Telescope Compact Array (ATCA)[1] between November 2010 to May 2016 (Project ID C2186; PI: A. Titmarsh). The ATCA is an array of six 22-m antennae located 500-km north-west of Sydney, at the Paul Wild Observatory. Five of these antennas can move along a 3 km-long east-west track or a 214-m north spur. The sixth antenna is fixed, being located 3 km from the track providing a maximum baseline length of 6 km along the east-west track. The ATCA is an Earth-rotation aperture synthesis radio interferometer and has 17 standard configurations that are designed to give optimum and minimum-redundancy coverage after a 12 hour observing period.

The complete data set consists of observations towards 323 6.7 GHz methanol masers sources between Galactic longitudes 310° and 20°, however, continuum data is only available for a subset of these sources due to limitations in the configurability of the correlator for observations made after August 2011. The observational setup included spectral windows targeting the $H_2O$ water maser, ammonia (1,1) and (2,2) transitions. The water maser results are published in Titmarsh et al. (2014, 2016), however, the ammonia and continuum data are currently unpublished. In this paper we use the unexploited continuum data towards 128 fields and covering 141 MMB sources. The data have been extracted from the Australia Telescope Online Archive (ATOA)[2].

### 2.2 Summary of observations

The array was set up in with various configurations, utilizing five of the six antennae in a compact orientation (see Table 2 for details). The sixth antenna was not used for this study due to the large gap in *uv*-coverage. The Compact Array Broad-band Backend (CABB; Wilson et al. 2011) simultaneously measures both low spectral resolution, large bandwidth radio continuum and many high spectral resolution, small bandwidth spectral line windows (zoom bands). The CABB was configured with 2 × 2 GHz continuum bands with 32 × 64 MHz channels to observe at 22.235 and 23.708 GHz, which covered the water maser (22 GHz) and ammonia (1,1) and (2,2) transitions. Therefore, our continuum bandwidth ranged from 21.223 to 24.708 GHz. Two zoom bands were used to obtain high resolution spectral data.

The target sources were divided into groups of 6 sources selected by their positions. To correct for fluctuations in the phase and amplitude caused by atmospheric effects, each group was sandwiched between observations of a nearby phase calibrator. The primary flux and bandpass calibrator 1934−638 was observed once during each set of observations to allow absolute calibration of the flux density scale. Each methanol maser position was observed at least four times over a six-hour period to ensure sufficient *uv*-coverage, giving a total on-source integration time of approximately 6 minutes (see Titmarsh et al. 2014, 2016 for details).

Continuum data are available for 133 fields, however, a small number of these overlap (pointing centres within 1 arcmin). The data for fields that overlap have been combined to improve *uv*-coverage

---

[1] The Australia Telescope Compact Array is part of the Australia Telescope National Facility (https://www.atnf.csiro.au/) which is funded by the Australian Government for operation as a National Facility managed by CSIRO.
[2] https://atoa.atnf.csiro.au

4    *A.L Patel et al.*

**Table 3.** Observational field parameters from all observed fields. The field names and pointing centres of the observations are based on the MMB names and positions.

| Field name | RA (J2000) (h:m:s) | Dec. (J2000) (d:m:s) | Field rms (mJy) | Beam size Major (arcsec) | Minor (arcsec) | Position angle (°) |
|---|---|---|---|---|---|---|
| G002.143+00.009 | 17:50:36.0 | -27:05:47 | 0.49 | 6.7 | 11 | 72 |
| G002.521−00.220 | 17:52:50.0 | -27:05:26 | 0.31 | 6.7 | 11 | 72 |
| G002.536+00.198 | 17:50:46.5 | -26:39:50 | 0.31 | 6.7 | 11 | 71 |
| G002.591−00.029 | 17:51:42.5 | -26:44:33 | 0.35 | 6.7 | 11 | 71 |
| G002.615+00.134 | 17:51:12.2 | -26:37:40 | 21 | 6.7 | 12 | 71 |
| G002.703+00.040 | 17:51:46.5 | -26:36:23 | 0.33 | 6.6 | 12 | 71 |
| G003.253+00.018 | 17:53:06.3 | -26:08:29 | 0.78 | 6.6 | 11 | 72 |
| G003.312−00.399 | 17:54:50.7 | -26:18:23 | 0.39 | 6.6 | 11 | 72 |
| G003.442−00.348 | 17:54:55.7 | -26:09:50 | 9.3 | 6.6 | 11 | 72 |
| G003.502−00.200 | 17:53:38.5 | -26:55:53 | 0.39 | 6.6 | 11 | 72 |

Note: only a small portion of the data is provided here, the full table is only available in electronic form.

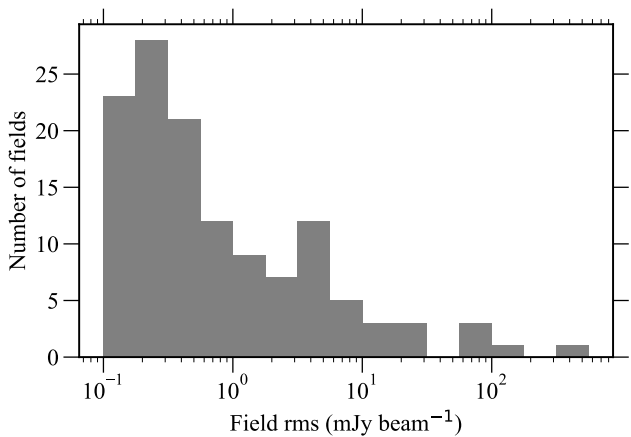

**Figure 1.** Histogram of the rms noise in the reduced radio maps. The histogram uses a bin size of 0.25 dex.

and sensitivity of the final images. Some of these fields were observed in multiple observing sessions and in these cases the data were also combined and imaged together. As a result of the various antennae configurations, limited $uv$-coverage and integration time, the largest well-imaged structures range from ∼40 arcsec. The total number of distinct fields imaged is 128, covering the positions of 141 methanol masers (see Table 3 for field parameters).

### 2.3 Data reduction and imaging

The calibration and reduction of these data were performed using the `MIRIAD` analysis package (Sault et al. 1995), following standard ATCA procedures. The data base was split into subsets containing the visibilites of the bandpass and phase calibrators and the target sources. For each subset, we performed a "flagging" procedure to eliminate any radio-frequency interference (RFI) and poor data. This also highlighted problems with the calibration and helped remove data affected by instrumental errors. The procedure was run iteratively until all poor data were eliminated from the visibility plots. We removed the frequency channels covering the data for the ammonia (1,1) and (2,2) the water maser transition to eliminate any potential contamination to the radio continuum from these spectral lines.

The primary flux and phase calibration were then performed and the calibration tables were copied over to the target sources. We have combined both continuum frequency bands to improve the SNR for the detected sources; this gives a total frequency coverage of ∼3.5 GHz. The calibrated data were subsequently cleaned and imaged using the `MIRIAD` tasks `INVERT`, `CLEAN` and `RESTOR`. The maps were deconvolved using a robust weighting of 0.5 and we use ∼ 200 cleaning components or until the first negative component was encountered. On all occasions the cleaning converges after a negative component was encountered. Additionally, we carried out the primary beam correction and a region the FWHM size of the primary beam was imaged using a pixel size of 1 arcsec, with 126 pixels along each side, which resulted in a image size of 2.1 × 2.1 arcmin. In Figure 1 we present a histogram of the number of fields as a function of their rms noise. The rms noise values were estimated from emission free regions close to the centre of the reduced maps (the noise values for individual maps are given in fourth column of Table 3).

### 2.4 Source extraction and final catalogue

The final reduced maps were examined for compact, high surface brightness sources using a nominal $3\sigma$ detection threshold, where $\sigma$ refers to the image rms noise level. The sources that had a peak flux density above the detection threshold were visually inspected in order to confirm the presence of a radio source and identify and exclude imaging artefacts from over-resolved sources and sidelobes.

In total, we have found 68 high frequency radio detections located within the 128 fields. We have created unique names for these detections based on their Galactic position. In Figure 2 we present the distribution of the peak flux density for these detections. We have identified five sources that display imaging artefacts due to poor $uv$-coverage; this can be seen visually as the maps contain large undulations. These sources are included in our final catalogue but we have added a note to indicate that the source parameters are less reliable. Eight fields were found to be associated with two radio sources, possibly indicating a degree of clustering. In three of these eight fields the radio sources are too close to separate and in these cases the emission has been classified as multi-peaked.

We use the `MIRIAD` task `IMFIT` to determine the positions, peak flux densities and sizes of the 68 detections. This was achieved by carefully drawing a polygon around the detection as an input to `IMFIT` that uses this to fit a two-component Gaussian around the emission. We were unable to fit sources G331.710+0.6041 and G332.290−0.091 with `IMFIT` due to the amount of undulation and

**Table 4.** Extracted source parameters from our detected radio sources. Column 1 is the radio name, which is a combination of the Galactic longitudes and latitudes determined by fitting ellipsoidal Gaussian using `IMFIT`, Columns 2 and 3 are the Right Ascension and Declination of each source, respectively. Columns 4 and 5 are the peak flux density, integrated flux density and errors, respectively and Column 6 are the observed sources sizes in arcsec.

| Radio name | RA (J2000) (h:m:s) | Dec. (J2000) (d:m:s) | $f_{peak}$ 23.7 GHz (mJy beam$^{-1}$) | $\Delta f_{peak}$ 23.7 GHz (mJy beam$^{-1}$) | $f_{int}$ 23.7 GHz (mJy) | $\Delta f_{int}$ 23.7 GHz (mJy) | Source size Major (arcsec) | Source size Minor (arcsec) | Source size Position angle (°) |
|---|---|---|---|---|---|---|---|---|---|
| G002.143+0.009 | 17:50:36.0 | -27:05:47 | 6.6 | 0.29 | 8.4 | 0.52 | 15 | 7.8 | 63 |
| G002.614+0.134 | 17:51:12.2 | -26:37:40 | 170 | 7.8 | 230 | 15 | 16 | 8.2 | 69 |
| G003.307−0.403 | 17:54:50.5 | -26:18:15 | 2.9 | 0.3 | 3.7 | 0.55 | 15 | 8 | 65 |
| G003.438−0.349 | 17:54:55.7 | -26:09:47 | 94 | 2.8 | 100 | 4.4 | 14 | 6.9 | 67 |
| G003.910+0.001 | 17:54:38.8 | -25:34:44 | 93 | 3.1 | 100 | 4.9 | 15 | 6.7 | 66 |
| G004.680+0.277 | 17:55:18.8 | -24:46:29 | 9.8 | 0.31 | 10 | 0.47 | 14 | 6.8 | 67 |
| G005.629−0.291 | 17:59:33.8 | -24:14:20 | 2.9 | 0.16 | 8 | 0.64 | 19 | 13 | 77 |
| G005.885−0.392 | 18:00:30.4 | -24:04:01 | 7900 | 230 | 8400 | 350 | 14 | 7.1 | 71 |
| G010.629−0.338 | 18:10:19.1 | -19:54:10 | 36 | 2.4 | 46 | 4.4 | 11 | 7.9 | -90 |
| G011.904−0.141 | 18:12:11.2 | -18:41:29 | 77 | 2 | 47 | 4.1 | 14 | 7.3 | -89 |
| G011.937−0.616 | 18:14:01.1 | -18:53:23 | 660 | 34 | 930 | 68 | 11 | 8.1 | 72 |
| G012.112−0.127 | 18:12:33.4 | -18:30:06 | 0.68 | 0.056 | 0.86 | 0.11 | 13 | 7 | -76 |

Note: only a small portion of the data is provided here, the full table is only available in electronic form.

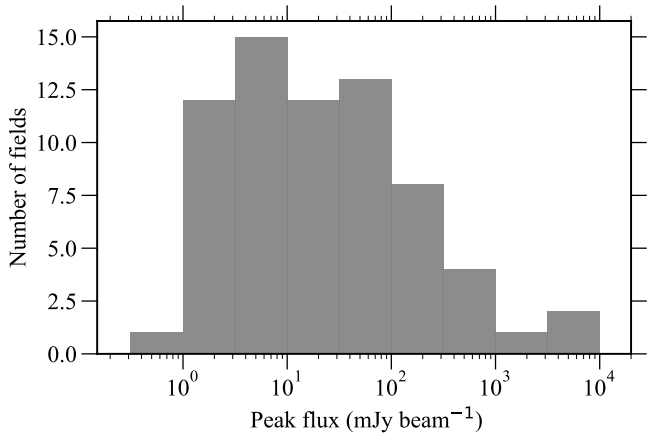

**Figure 2.** Histogram of the peak flux density of the detected radio sources. The histogram uses a bin size of 0.50 dex.

the shape of the sources; we use `SExtractor` to manually fit a two-component Gaussian around the radio emission for these two sources.

## 3 RESULTS

### 3.1 Initial inspection and classification

Our final catalogue presents the results of observations towards 141 MMB methanol masers that resulted in the detections of 68 high-frequency radio sources at 23.7 GHz. In Table 4 we present the radio name, RA and Dec, peak and integrated fluxes and source major, minor axes and position angle for our detections. Radio emission can result from different astrophysical processes and to help identify the origin of the emission, we have produced maps of the fields with high-frequency radio emission and overlay the positions of the following star-formation: methanol masers from the MMB survey (Caswell 2010; Green et al. 2010; Caswell et al. 2011) and 5 GHz radio sources taken from the literature (see Table 5 for more detail; Becker et al. 1994; Urquhart et al. 2007, 2009; Purcell et al. 2013; Irabor et al. 2023).

**Table 5.** Observational parameters for the archival 5 GHz catalogues used to compare and calculate the spectral index measurements.

| Catalogue | Synthesised beam (arcsec) | Sensitivity rms (mJy beam$^{-1}$) | Largest angular scale (arcsec) | Reference |
|---|---|---|---|---|
| Magpis | 5 | 0.30 | 100 | (1) |
| CORNISH−North | 1.5 | 0.43 | 14 | (2) |
| CORNISH−South | 2.5 | 0.11 | 40 | (3) |
| RMS | 1.5-2.5 | 0.32 | 20 | (4) |

References. (1) Becker et al. (1994), (2) Purcell et al. (2013), (3) Irabor et al. (2023), (4) Urquhart et al. (2007).

We visually inspected these maps to determine whether our high-frequency radio sources are positionally associated (within $3\sigma$; See Fig. 7) with these star-formation tracers. Figure 4 presents a Venn diagram showing the positional association between methanol masers, low-frequency radio and dust emission within the detected high-frequency radio ellipse. Of the 68 high-frequency radio sources detected we have found that 14 are associated with dust emission and methanol maser(s) but not with 5 GHz radio emission. Three of our detections lie outside the latitude limit for the surveys considered and so the lack of a 5 GHz counterpart is not significant but for the other 11 the lack of a low frequency counterpart suggests they are likely to be in a very early stage in their evolution and therefore good HC H II regions candidates; we will discuss the nature of these sources in more detail in Sect. 5. We find 5 GHz counterparts for 25 of our high-frequency detections and we expect these to be more evolved and we classify them as UC H II regions candidates. We also find 15 high-frequency radio sources that are not associated with a methanol maser but are associated with dust and 5 GHz radio emission and these are likely to be more evolved H II regions for which the conditions necessary for methanol maser excitation have been disrupted due to the expansion of the H II region.

The H II regions discussed so far account for 54 of the 68 radio detections, leaving the nature of the remaining 14 to be investigated. There is one 23.7 GHz radio source (G019.491+0.135) with a low frequency radio counterpart that is not associated with any dust or methanol maser emission. The lack of any associated dust or mid-infrared emission with this source and the steep negative spectral index ($\alpha = -1.33 \pm 0.06$; see Section 3.2.3 for details) is consistent with it being an extragalactic radio source.

6   *A.L Patel et al.*

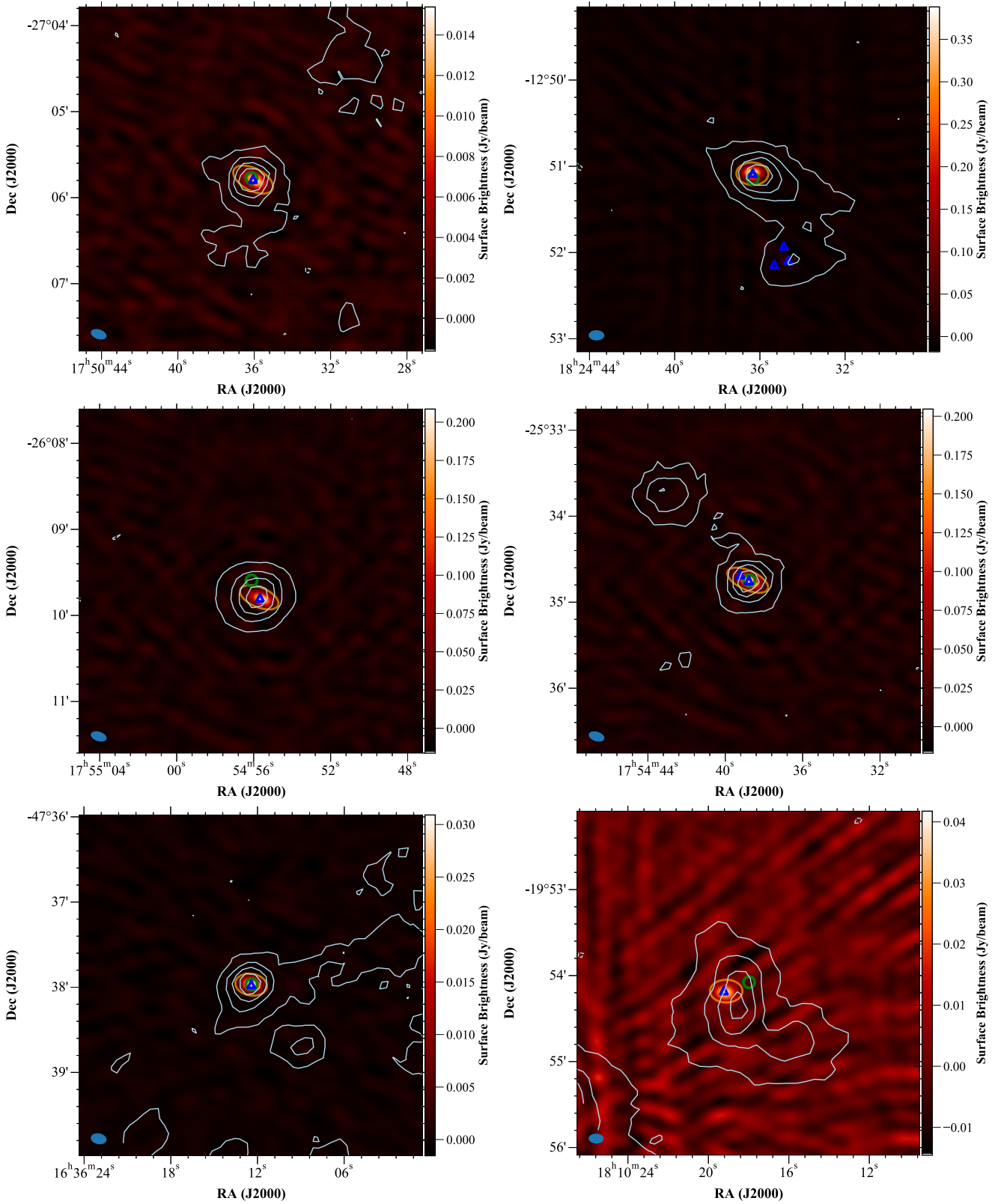

**Figure 3.** Six examples of high-frequency radio detections. The orange ellipse shows the resultant fit to the radio emission while the green circles shows the position of the methanol masers located in the field. The blue triangles show the position of any low-frequency (5 GHz) radio counterparts. The grey contours trace the 870 μm dust emission from ATLASGAL (Schuller et al. 2009). The filled blue ellipse in the bottom left hand corner of each image indicates the size and orientation of the synthesised beam.

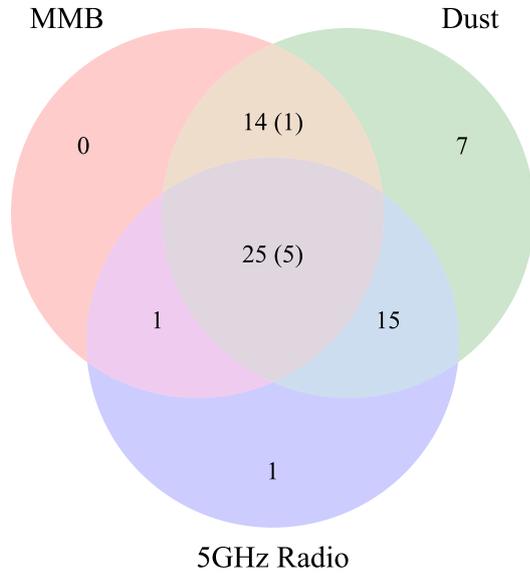

**Figure 4.** Venn diagram presenting the total number of radio sources and the corresponding associated methanol masers, ATLASGAL dust clumps and 5 GHz radio sources. Five of the high-frequency radio sources do not have a star forming counterpart. The values in brackets show the number of water masers that are associated with a 6.7 GHz methanol maser.

We find seven high-frequency radio sources that are only associated with dust; it is difficult to assess the nature of these objects from their association with dust alone and so we have examined their mid-infrared environments by creating three-colour composite maps made from data in the 4.5, 5.8 and 8.0 μm IRAC (Fazio et al. 2004) bands from the Galactic Legacy Infrared Mid-Plane Survey Extraordinaire (GLIMPSE) survey (Benjamin et al. 2003); these images are presented in Figure 5.

Visual inspections of these images have found two radio sources (G012.632−0.017 and G014.986−0.122) that are coincident with extended mid-infrared emission, and one radio source (G014.599+0.020) that is coincident with compact mid-infrared emission. We have classified the former as being H II regions and the latter as an UC H II region; all three are shown in panels a, b and c of Figure 5, respectively. A further two radio sources (G331.334−0.335, G332.644−0.607) are coincident with the inner edge of large mid-infrared bubbles and so we classify these objects as photodissociation regions (PDRs; Genzel et al. 1989; these are shown in panels d and e of Figure 5 respectively). One of the radio sources (G003.307−0.403) has no mid-infrared association and is offset from the peak of the dust emission and methanol maser; it is likely that this source is a background extragalactic radio source (see panel f of Figure 5). We find another radio source (G332.098−0.419) in this category which is slightly offset from a bright nearby mid-infrared source that is saturated towards its centre. The nature of this radio source is not clear from the evidence we have available, but SIMBAD classifies this source as a radio star (see panel g of Figure 5).

We have identified one source (G331.542−0.067) that is only associated with a methanol maser and low-frequency radio emission (Figure 5h). The high-frequency radio detection is associated with an extended and more evolved H II region whose emission is coincident with the edges of the ionized gas, suggesting this source is a combination of a PDR and an H II region. Although the radio emission is not associated with the nearby ATLASGAL dust clump AGAL331.546−00.067 (Contreras et al. 2013), inspection of the dust map reveals the presence of low surface brightness extended 870 μm emission towards the position of the radio emission. The correlation between the radio and MMB positions and the presence of dust emission is consistent with this being classified as an H II region.

Five of our high-frequency radio detections have no associations with any star-forming tracers mentioned so far. These could either be planetary nebulae (PNe) or extragalactic radio sources. If they are PNe, we should expect them to be embedded in dust and appear as mid-infrared point sources; however, since these five do, not we conclude they are more likely extragalactic in origin.

In total, we have identified 58 H II regions, two radio sources associated with PDRs, seven extragalactic sources and one possible radio star candidate. We excluded the extragalactic source, the PDRs and the radio star from further analysis.

## 4 CONFIRMING THE NATURE OF H II REGION CANDIDATES

In the previous section we identified 58 embedded H II region candidates. In this section we investigate their correlation with other surveys, and where possible, calculate the spectral index to determine the nature of these high-frequency radio detections to identify a reliable sample of HC H II region candidates.

### 4.1 Comparison with ATLASGAL classifications

We compare these H II region candidates with classifications made by the ATLASGAL team (Urquhart et al. 2014, 2018, 2022) to confirm our classification and identify previously unknown H II regions aiming to identify potentially optically thick objects. In Figure 6 we show the distribution of the ATLASGAL classifications of the host dust clumps.

The majority of the clumps are found to be associated with H II regions (37 corresponding to 64.3 per cent by the ATLASGAL team), which is consistent with our initial interpretation. The most interesting in terms of identifying very young H II regions are those clumps previously classified as protostellar (three corresponding to 5.4 per cent) and YSOs (eight corresponding to 14.3 per cent) as they have not previously exhibited any traits attributed to UC H II regions and are likely to be very young. In total, 37 of our H II region candidates have been confirmed by the ATLASGAL to be compact or UC H II regions and we have have identified eleven as being even younger.

Of the remaining nine radio sources we have classified as H II regions, six had been classified as PDRs by the ATLASGAL team, and three as being associated with a complicated mid-infrared environment that makes a definitive classification unreliable. We have inspected the mid-infrared environments of these objects and can confirm that the radio emission is associated with more complicated and larger-scale structures. Since the radio emission from these objects is not associated with H II regions, we exclude them from further analysis. This leaves us with 49 H II region candidates including the one H II region that is associated with only low-frequency radio and methanol-maser emission.

8   *A.L Patel et al.*

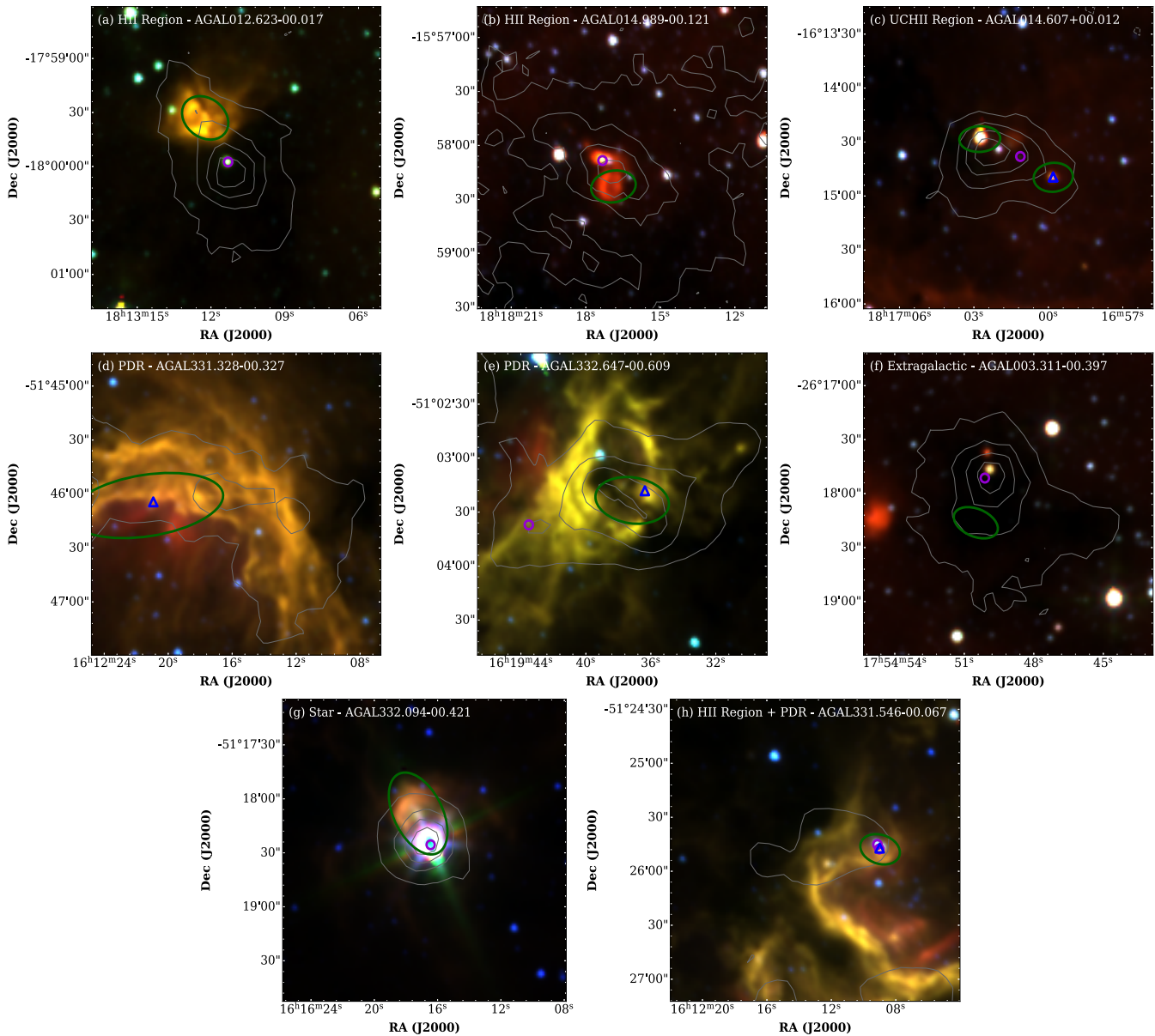

**Figure 5.** Three colour images of the seven radio sources only associated with dust emission. We utilise the 4.5, 5.8 & 8.0 μm IRAC bands from the GLIMPSE survey. The green ellipse shows the position of the high-frequency (23.7 GHz) radio source and the purple circles shows the position of the methanol maser. The blue triangles show the position of the low-frequency (5 GHz) radio counterpart, the grey contours trace 870 μm dust emission from ATLASGAL.

### 4.2 Angular offset distribution

In the previous section, we have matched our high-frequency radio sources with the positions of the different star-forming tracers by eye from the radio emission maps and three-colour mid-infrared composite images (e.g., see Figure 3 and 5) using the fitted ellipse to the radio emission as a guide. In Figure 7 we present histogram plots of the angular offset distribution from the three star-formation tracers previously discussed. The distributions shown in these plots have an approximate Gaussian profile and a high density of matches at very small radial offsets, dropping sharply for the 23.7 GHz−MMB and 23.7 GHz−5 GHz matches while the Radio−ATLASGAL falls less sharply due to the more extended distribution of dense molecular material within a cloud. We use a $3\sigma$ threshold to distinguish between real and chance associations; these are shown on the plots shown in Figure 3 and clearly shows that the majority of the associations are robust.

We find one radio−ATLASGAL match that has an angular offset greater than 18 arcsec; however, this is due to the radio emission being positioned on the outer edge of the dust clump. The radio source G336.990−0.025 is extended in the mid-infrared and appears to be part of a larger structure that hosts a 5 GHz radio counterpart. Additionally, the lack of a methanol maser counterpart would suggest that this HII region is more evolved and is excluded from further analysis. Of the 23.7 GHz−MMB matches, two matches have an angular offset greater than 8 arcsec (i.e. $> 3\sigma$). In both of these cases, the high-frequency radio emission is extended and the methanol maser is positioned on the edge of the radio emission. Despite the large offset, these sources are still associated with the methanol maser and are still considered to be HII regions. There is one 23.7 GHz–5 GHz match that has an angular offset greater than 3 arcsec (i.e. $> 3\sigma$). In this case there are two 5 GHz radio sources (3.910+0.001 and 3.912+0.000 offset: 0.6 and 7.9 arcsec, respectively; Becker et al. 1994) within the high frequency radio ellipse. Discarding the radio match with the larger offset we find the positional correlation between the radio matches to be very strong with all matches being within 3 arcsec ($\sim 3\sigma$).

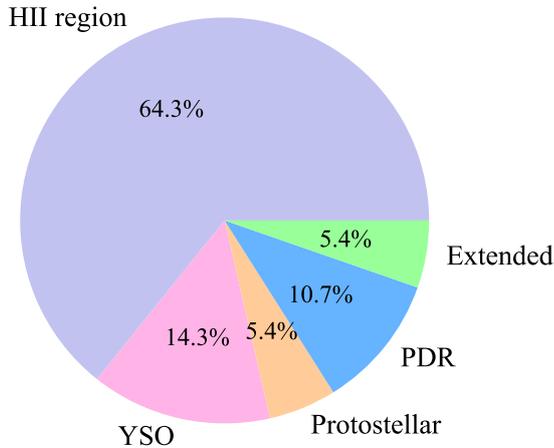

**Figure 6.** Distribution of the ATLASGAL classification for the clumps associated with 57 high-frequency radio classified as H II region candidates in Sect. 3.

### 4.3 Water maser emission

Water masers are collisionally pumped and tend to occur at the location where the protostar interacts with the surrounding environment, such as the interface between the outflow and the ambient gas (Walsh et al. 2011). Titmarsh et al. (2014, 2016) searched the environments of 323 MMB methanol masers for water maser emission and found an association rate of $\sim 48$ per cent using a 3 arcsec offset criterion. We investigate the correlation between the detected water masers and our radio continuum sources, where we apply the same angular offset as for the 23.7 GHz–MMB associations (see middle panel of Fig. 7). We find that of the 48 radio sources associated with a methanol maser and classified as an H II region, six show water maser emission within 8 arcsec.

Five of these water masers are found to be associated with a methanol maser, 870 μm dust emission and 5 GHz radio emission, whereas one source is only coincident with methanol maser and 870 μm dust emission. This corresponds to a $\sim 12$ per cent association rate between water masers, methanol masers, and radio continuum emission. The incidence of water maser emission is significantly lower for methanol maser sources associated with radio emission compared to the water masers that are coincident with just methanol masers alone. This indicates that as a H II region forms around a MYSO the water maser activity decreases, possibly due to disruption of the circumstellar envelope by the expanding bubble. Given this, water masers most likely to trace earlier stages of both low- and high mass star formation (Deacon et al. 2007; Urquhart et al. 2022). We find that two of our HC H II region candidates present both water and methanol masers emission (G005.885−0.392 and G018.735−0.227); we discuss these further in Section 5.

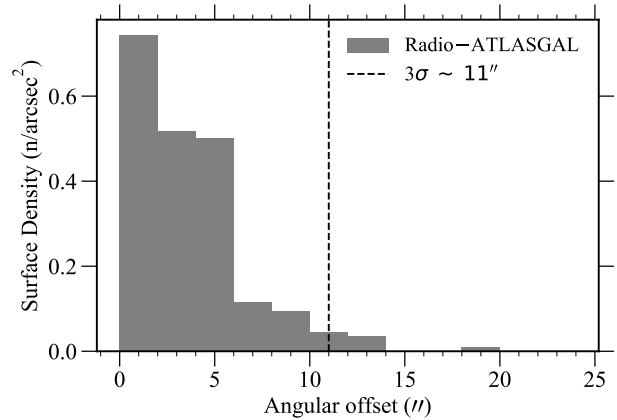
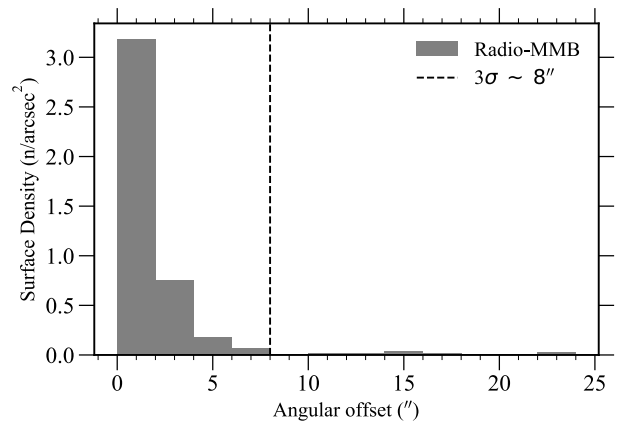
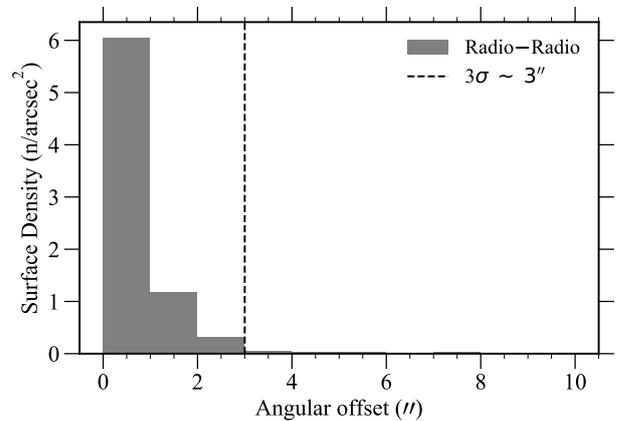

**Figure 7.** Histograms of the angular separation between detected radio sources at 23.7 GHz and their nearest ATLASGAL dust clump (top panel), MMB maser (middle panel), and 5 GHz radio counterpart (lower panel). The first and middle panel are binned using a value of 2 arcsec and the lower panel is binned using a value of 1 arcsec. The dotted line shows the $3\sigma$ radial cutoff and the approximate value is given in the legend.

### 4.4 Spectral Index

From the section above we have identified 48 H II regions, 32 of which are associated with both 23.7 GHz and 5 GHz radio emission. We expect UC H II to have a turnover frequency at $\sim 5$ GHz (Beltrán et al. 2007; Keto & Klaassen 2008; Hoare et al. 2007) and so we would expect these to approach a flat spectrum (of $\alpha \simeq -0.1$)

10   *A.L Patel et al.*

between these frequencies, while HC H II regions are expected to be optically thick and so they should have a positive slope between them $\alpha \simeq 1-2$ (Kurtz 2005). The spectral index is therefore a useful way to distinguish between optically thin and thick H II regions.

We can measure the spectral index of these sources using the following equation:

$$\alpha = \frac{\ln(S_{23.7\,\text{GHz}}/S_{5\,\text{GHz}})}{\ln(23.7/5)} \qquad (1)$$

where $S_\nu$ is the peak flux density at the given frequency, in our case 23.7 and 5 GHz. The uncertainty of our spectral index determination is:

$$\Delta_\alpha = \frac{\sqrt{(\sigma_{23.7\,\text{GHz}}/S_{23.7\,\text{GHz}})^2 + (\sigma_{5\,\text{GHz}}/S_{5\,\text{GHz}})^2}}{\ln(23.7/5)} \qquad (2)$$

where $\sigma_\nu$ is the error of the flux density at the given frequency. The spectral index can only be determined when, at both frequencies, flux densities and the source is compact.

For our sample of 5 GHz radio sources we compare the integrated fluxes against the peak flux to determine what we call the y-factor (i.e., $f_{\text{int}}/f_{\text{peak}}$). To keep spatial filtering to a minimum we limit the y-factor < 2, to ensure that our sample of 5 GHz radio sources are unresolved or compact. We also utilise data measured in the IRAC wavebands to explore the mid-infrared environments of the embedded radio emission. We pay particular attention to the clumps that appear extended or fluffy as they are likely to be significantly affected by spatial filtering. Based on the available evidence we find 15 are extended and we excluded these from our spectral index analysis. This leaves us with a sample of 19 H II regions that satisfy both conditions.

The largest spatial scales to which the 5 GHz surveys are sensitive to is ~15-100 arcsec and so it is comparable to the 23.7 GHz observations(~ 20 arcsec), however, to account for the difference in beam sizes we use the peak flux density at 23.7 GHz ($\theta_{\text{beam}} \sim 10''$) and the integrated flux at 5 GHz ($\theta_{\text{beam}} < 5''$) for all compact objects. Three 23.7 GHz radio detections have multiple 5 GHz counterpart within the 23.7 GHz beam; in these cases we sum the integrated fluxes of the 5 GHz sources to determine the flux used to calculate the spectral index.

As a consistency check we extracted the 5 GHz radio maps for all the radio sources covered by multiple surveys and measured the flux densities and uncertainties directly from the maps. We found these measurements were within 10 per cent of the published catalogues and have therefore used the parameters from the catalogues for completeness. We also compared the fluxes for 7 sources that have a counterpart in both the CORNISH and MAGPIS surveys and find these are in good agreement. Given the difference between spatial scales of these two surveys and agreement in their flux values it is clear that differential spatial filtering at 5 GHz does not have a significant effect on the derived spectral indices. We present the 5 GHz radio maps for the 19 radio detections and 9 non-detection in Appendix B[3].

It is also possible to determine a lower limit to the spectral index $\alpha_{\text{min}}$ for our 23.7 GHz detections by using the upper limit for

---

[3] Supplementary online material

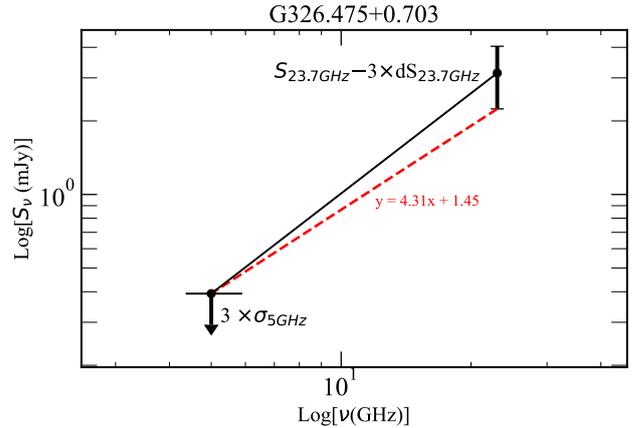

**Figure 8.** Flux density measurement as a function of frequency for G326.475+0.703 showing the measured 23.7 GHz flux and the 5 GHz upper limit determined from the $3\sigma$ rms noise from the image. The error bar on the 23.7 GHz flux is $3\sigma$. The red line shows the result of a linear fit to the 5 GHz upper limit and the 23.7 GHz flux minus, 3 times the uncertainty on the 23.7 GHz flux measurement. The slope represents the lower limit to the spectral index of this source.

the flux density at the lower frequency. We use a formalism from Yang et al. 2019:

$$\alpha_{\text{min}} = \frac{\ln[(S_{23.7\,\text{GHz}} - 3 \times \text{d}S_{23.7\,\text{GHz}})/(3 \times \text{d}S_{5\,\text{GHz}})]}{\ln(23.7/5)} \qquad (3)$$

where $\text{d}S_\nu$ is the uncertainty in the flux density at the given frequency. Here, $\text{d}S_{23.7\,\text{GHz}}$ is determined from emission free regions of the maps, whereas $\text{d}S_{5\,\text{GHz}}$ is the measured value towards the position of the 5 GHz radio source that have been extracted from the CORNISH 5 GHz database (Purcell et al. 2013)[4], RMS 4.8 GHz database (Lumsden et al. 2013)[5] and CORNISH-South survey (Irabor et al. 2023). This equation assumes a $3\sigma$ detection threshold from the 5 GHz radio maps and subtracts $3\sigma$ uncertainty from the measured 23.7 GHz flux density and ensures we are determining an accurate lower limit. In Figure 8 we show the calculation of the lower limit for the spectral index for one of the 5 GHz non-detections.

In total, we have obtained spectral index measurement for 28 compact H II regions (19 reliable measurements and 9 lower limits; see Table 6 for detail). Figure 9 shows the distribution of the number of sources as a function of their spectral index, $\alpha$ and the lower limit of the spectral index, $\alpha_{\text{min}}$. We notice that there is a peak centred at 0, indicating that the majority of these H II regions are optically thin. In this study, we consider detections that are mid-infrared point sources or mid-infared quiet and those that have a positive spectral index to be optically thick between 5 and 23.7 GHz. All but seven sources have a flat or positive spectral index as we would expect for H II regions. 13 radio detections have a positive spectral index or lower limit constraint suggesting the emission could be optically thick between 5 and 23.7 GHz; consistent with them being HC H II region candidates. We will discuss these in more detail in Section 5.

**Table 6.** Spectral index and lower limit measurements for the 19 high frequency radio detections that present 5 GHz radio emission. The average offset and size has been taken for sources that contain more than one 5 GHz radio counterpart and is represented with an †. The 5 GHz integrated flux upper limits are given as $3\sigma$ thresholds.

| Radio name | | $f_{int}$ 5 GHz (mJy) | $\Delta f_{int}$ 5 GHz (mJy) | Observed 5 GHz size (arcsec) | $f_{peak}$ 23.7 GHz (mJy beam$^{-1}$) | $\Delta f_{peak}$ 23.7 GHz (mJy beam$^{-1}$) | Offset (arcsec) | | $\alpha$ | $\Delta\alpha$ | Reference catalog |
|---|---|---|---|---|---|---|---|---|---|---|---|
| G002.143+0.009 | | 5.7 | 0.2 | 1.5 | 6.6 | 0.29 | 0.41 | | 0.096 | 0.023 | (1) |
| G002.614+0.134† | | 120 | 0.78 | 7.0 | 170 | 7.8 | 1.7 | | 0.21 | 0.019 | (1) |
| G003.438−0.349 | | 94 | 0.41 | 2.3 | 94 | 2.8 | 0.67 | | 0.00054 | 0.013 | (1) |
| G003.910+0.001† | | 37 | 0.34 | 1.9 | 93 | 3.1 | 3.7 | | 0.58 | 0.014 | (1) |
| G004.680+0.277 | | 5.4 | 0.54 | 5.2 | 9.8 | 0.31 | 1 | | 0.39 | 0.043 | (1) |
| G005.885−0.392 | | 2200 | 1.4 | 3.3 | 7900 | 230 | 0.56 | | 0.81 | 0.012 | (1) |
| G011.904−0.141† | | 42 | 1.8 | 3.6 | 76 | 2.0 | 1.6 | | 0.38 | 0.02 | (2) |
| G012.112−0.127 | < | 1.2 | | | 0.68 | 0.06 | | > | −0.96 | | (2) |
| G014.229−0.510 | < | 1.0 | | | 1.5 | 0.34 | | > | −0.31 | | (2) |
| G014.599+0.020 | | 4.4 | 1.1 | 2.1 | 6.6 | 0.12 | 0.17 | | 0.26 | 0.1 | (2) |
| G014.610+0.012 | < | 1.0 | | | 6.2 | 0.27 | | > | 0.86 | | (2) |
| G018.461−0.004 | | 340 | 32 | 3.1 | 420 | 14 | 0.5 | | 0.13 | 0.04 | (1) |
| G018.665+0.029 | | 5.7 | 0.85 | 2.0 | 5.5 | 0.49 | 0.046 | | −0.014 | 0.072 | (2) |
| G326.475+0.703 | < | 0.39 | | | 3.1 | 0.3 | | > | 1.2 | | (4) |
| G327.402+0.445 | | 87 | 0.31 | 1.1 | 120 | 2.7 | 0.45 | | 0.23 | 0.0091 | (3) |
| G331.442−0.187 | | 62 | 5.4 | 13 | 69 | 1.3 | 0.36 | | 0.075 | 0.037 | (4) |
| G331.542−0.067 | | 75 | 0.55 | 0.43 | 330 | 46 | 2.2 | | 0.95 | 0.058 | (3) |
| G331.710+0.604 | < | 0.6 | | | 1.5 | 0.12 | | > | 0.38 | | (4) |
| G332.826−0.549 | | 1300 | 4.5 | 3.2 | 4200 | 180 | 0.48 | | 0.78 | 0.018 | (3) |
| G332.987−0.487 | < | 0.36 | | | 8.0 | 1.3 | | > | 1.9 | | (4) |
| G333.029−0.063 | | 5.6 | 0.15 | 2.6 | 31 | 1.3 | 0.57 | | 1.1 | 0.02 | (4) |
| G336.983−0.183 | | 18 | 0.25 | 0.42 | 56 | 1.9 | 1.1 | | 0.73 | 0.015 | (3) |
| G338.280+0.542 | < | 0.38 | | | 3.8 | 0.24 | | > | 1.3 | | (4) |
| G338.566+0.110 | | 5 | 0.19 | 2.8 | 8.8 | 0.32 | 0.77 | | 0.36 | 0.022 | (4) |
| G338.922+0.624 | | 140 | 0.21 | 1.5 | 190 | 7.9 | 0.36 | | 0.17 | 0.017 | (3) |
| G338.925+0.556 | < | 0.38 | | | 12 | 1.1 | | > | 2.0 | | (4) |
| G339.622−0.121 | < | 0.42 | | | 4.9 | 0.25 | | > | 1.4 | | (4) |
| G339.980−0.539 | | 140 | 0.65 | 5.5 | 52 | 30 | 0.79 | | −0.64 | 0.24 | (4) |

References. (1) Becker et al. (1994), (2) Purcell et al. (2013), (3) Urquhart et al. (2007), (4) Irabor et al. (2023).

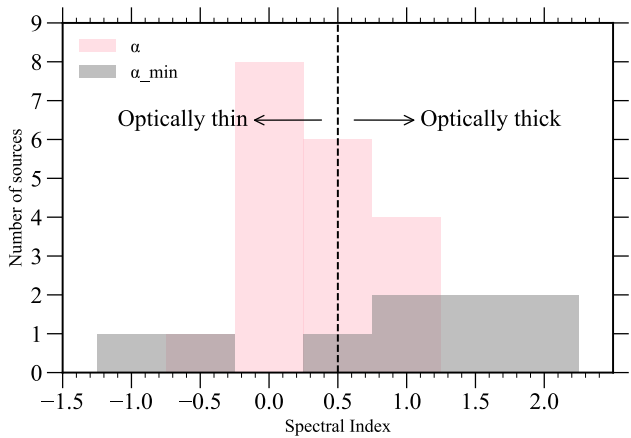

**Figure 9.** Histogram showing the number of fields observed as a function of the spectral index and lower limit in light grey and grey respectively. We have measured a reliable spectral index value for 19 sources and the lower limit for 9 sources. The data has been binned using a bin size of 0.50. The dotted line indicates a estimated threshold for distinguishing between optically thick and thin nebulae.

### 4.5 Physical properties

Our observations targeted 141 methanol masers that encompass 99 distinct ATLASGAL dust clumps from which, we have identified 49 H II regions. We have extracted physical properties such as bolometric luminosity, full-width-half-maximum (FWHM) mass and $L_{bol}/M_{clump}$ ratio from the ATLASGAL catalogue (Urquhart et al. 2022). In Figure 10 we show cumulative distribution functions for the FWHM clump mass, luminosity and luminosity-to-mass ratio for the full population of ATLASGAL dust clumps classified as being protostellar, YSO and H II region and our sample of 49 H II regions. We use the two sample Kolmogorov-Smirnov (KS) test to compare the similarities between our sample of H II regions against the different evolutionary types for the various derived parameters. The test measures the largest vertical distance between two distributions, which is referred to as the KS statistic ($D$). The confidence value derived from the KS statistic is referred to as the $p$-value. Our null hypothesis states that the distribution of our H II region sample is identical to the general population of embedded H II regions identified by ATLASGAL. This can be rejected if the $p$-value is lower than $3\sigma$ (i.e. < 0.0013). Inspection of the FWHM mass (upper panel of Fig. 10) between the evolutionary classifications and our sample of H II regions reveal no significant difference demonstrating the data is consistent with the null hypothesis (KS test returns a $p$-value > 0.41). Whereas, the distribution for luminosity and luminosity-to-mass ratio increases with respect to the evolutionary sequence (middle and lower panel of Fig. 10). For both of these parameters the *p*-value is < 0.0013 and so we are able to reject the null hypothesis and thus can confirm that the observed difference is statistically significant.

---

[4] https://cornish.leeds.ac.uk/public/index.php
[5] http://rms.leeds.ac.uk/cgi-bin/public/RMS_DATABASE.cgi

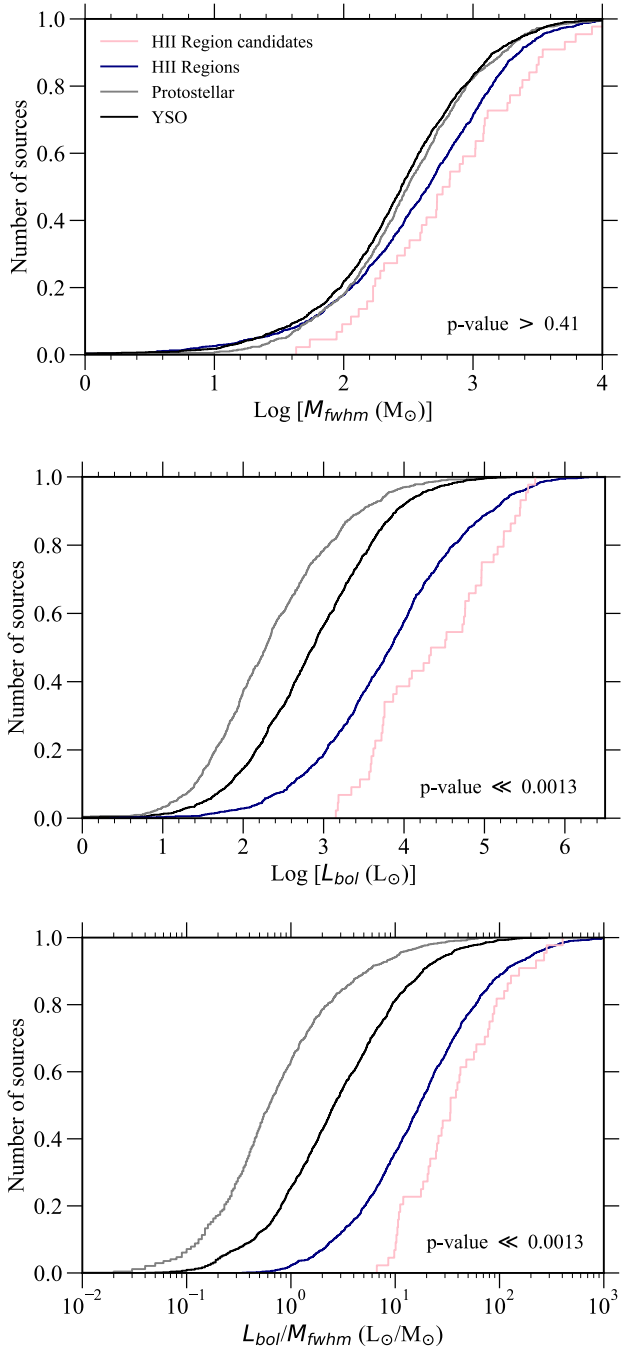

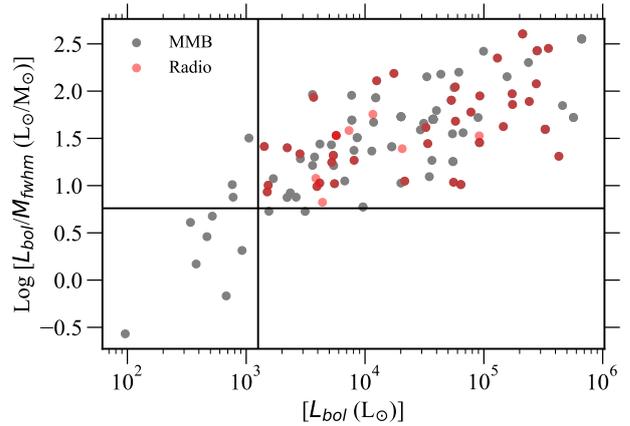

**Figure 11.** The bolometric luminosity for the full population of methanol masers (grey) and H II regions (red) as a function of their luminosity-to-mass ratio. There are ∼ 10 sources that are offset from the central MMB methanol maser and appear as pink data points.

**Figure 10.** Cumulative distribution functions for bolometric luminosity (top panel), FWHM Mass (middle panel) and $L_{bol}/M_{clump}$ ratio (lower panel) for the full population of ATLASGAL dust clumps classified as being protostellar, YSO and H II region. In pink we present our sample of 49 H II regions. The *p*-value generated from a KS test is represented in the bottom left hand corner of each panel.

In Figure 11 we present a scatter plot of the luminosity-to-mass ratio as a function of bolometric luminosity for all methanol masers observed as part of this study (grey) and the maser sources that have high frequency (23.7 GHz) radio emission (red). We have reliable measurements for 92 of the 99 ATLASGAL clumps and 44 of the 49 H II regions. Inspecting the plot shows that the H II regions are located in a region of the parameter space bounded by > $10^3\,L_\odot$ and a $L_{bol}/M_{clump}$ ratio of > $5.1\,L_\odot\,M_\odot^{-1}$. This luminosity is consistent with the presence of a 8-10 $M_\odot$ main sequence star that has evolved sufficiently to be able to form an H II region. Almost 90 per cent of the methanol masers are located in this part of the parameter space, which reinforces the results of previous studies that have concluded that methanol masers are exclusively associated with high-mass protostellar objects (e.g., Minier et al. 2003; Breen et al. 2013).

### 4.6 Lyman continuum flux vs Bolometric luminosity

As discussed in the introduction, radio jets and optically thick H II region have similar spectral indices. One way to discriminate between the two is by looking at the relationship between Lyman flux and bolometric luminosity. Radio jets have a lower power-law relationship between the Lyman photon flux and bolometric luminosity than H II regions and this can be used to distinguish between these two types of objects. A recent study by Purser et al. (2021) presents radio observations towards 56 massive star forming regions at different evolutionary stages. Thermal radio emission was detected from 94 per cent of their MYSO sample, of which 84 per cent present jet-like features. In addition 26 sources were classified as ionised jets in an earlier study (Purser et al. 2016). In order to confirm that our sources are H II regions and not MYSOs driving ionised jets we can compare luminosities and calculate the Lyman continuum output rate $N_i$ (Panagia 1973; Thompson 1984; Carpenter et al. 1990; Urquhart et al. 2013b), which is represented by Eqn 4,

$$\left(\frac{N_i}{\text{photon s}^{-1}}\right) = 8.9 \times 10^{43} \left(\frac{S_\nu}{\text{mJy}}\right) \left(\frac{D}{\text{kpc}}\right)^2 \left(\frac{\nu}{\text{GHz}}\right)^{0.1} \quad (4)$$

where $S_\nu$ is the integrated radio flux density at the measured frequency $\nu$, $D$ is the heliocentric distance to the source. This equation assumes that the H II regions are optically thin. In the case of HC H II regions that are likely to be optically thick below ∼ 40 GHz, the

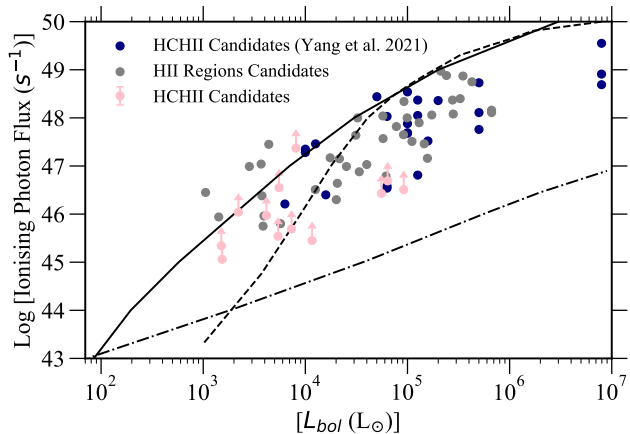

**Figure 12.** Lyman continuum photon flux as a function of bolometric luminosity for the H II region sample. The pink and grey circles show our sample of HC H II and H II regions respectively. The dashed line indicates the relationship for OB ZAMS stars models (Davies et al. 2011; Martins et al. 2005) and the dot-dashed line represents a fit to radio jets founded in Purser et al. 2016 and Purser et al. 2021. The solid black line corresponds to the Lyman continuum expected from a blackbody with the same radius and temperature as a ZAMS star (see Sánchez-Monge et al. 2013 for more detail.)

Lyman flux will be significantly underestimated (Urquhart et al. 2013b).

In Figure 12, we show the Lyman photon flux as a function of bolometric luminosity for the H II region sample. We include the luminosity–Lyman photon flux relationship for OB ZAMS stars from a table of values given in Davies et al. 2011 (also see Martins et al. 2005). The dot-dashed line represents the power law relation for the radio jets identified by Purser et al. (2016) where we use Eqn 4 to convert the radio luminosity into Lyman photon flux. For the HC H II candidates lower limits of the Lyman flux are shown as it is likely their flux densities are underestimated due to their optically thick nature. In our case, the sample of H II regions all present Lyman photon fluxes above those expected by radio jets suggesting these objects are genuine H II regions. Additionally, we include the 24 HC H II regions and candidates identified in Yang et al. 2021 (and references therein). Our sample of HC H II region candidates matches the known distribution of HC H II regions well. We also find that our candidates extend the luminosity range towards lower values by one order of magnitude, which suggests we are capable of identifying HC H II regions associated with less massive, high-mass stars.

Comparing the measured Lyman flux with that predicted by stellar atmosphere models of ZAMS stars in Fig. 12 reveals a significant divergence for H II regions with luminosities equivalent to early B-type stars with the measured value lying above the standard stellar atmosphere model for bolometric luminosities below $10^4$ L$_\odot$. This disagreement between the measured and predicted Lyman flux has been reported by a number of previous studies (e.g., Lumsden et al. 2013; Sánchez-Monge et al. 2013; Urquhart et al. 2013b). Sánchez-Monge et al. (2013) noted that there is much better agreement between the measured Lyman flues for early B-type stars and the expected Lyman continuum emission from a blackbody with the same radius and effective temperature as a ZAMS star (this is indicated on Fig. 12 by the solid curve). The origin of this excess has been studied by Cesaroni et al. (2016), who present two likely explanations. The first is known as the flashlight effect (e.g., Yorke & Bodenheimer 1999) where the majority of the stellar photons are thought to flow along the axis of a bipolar outflow, which results in an underestimation of $L_{bol}$. The second scenario assumes the Lyman excess is a due to the excess of UV photons from nearby stellar neighbours. Cesaroni et al. (2016) consider 200 target sources, 67 of which present Lyman continuum excess, however, they conclude that this excess is not easily understood.

## 5 DISCUSSION OF HC H II REGION CANDIDATES

In the previous section we have identified a number of high frequency radio sources associated with protostellar and YSO-hosting clumps as classified by the ATLASGAL team. We have also identified some sources classified as H II regions that have positive spectral index and so might also be very young and therefore HC H II region candidates. In this section we will discuss the evidence available for these sources in an effort to compile a reliable sample of HC H II region candidates. To assist with this evaluation we have created three colour composite images from GLIMPSE and visually inspected the 22 μm WISE band (Wright et al. 2010) and the 70 μm Hi−GaL band (Molinari et al. 2010a) and conducted a search of SIMBAD to identify relevant counterpart within 1 arcmin of each source. Below we will review the available evidence.

### 5.1 Protostellar

**G326.475+0.703 (Figure 13a):** The detected radio emission is closely associated with the MMB source G326.475+00.703 (Green et al. 2012, which is offset by 0.8 arcsec. There is a second MMB source located 29.1 arcsec to the south east (G326.476+00.695; Green et al. 2012) that is associated with bright mid-infrared emission and has been previously classified as a YSO by the Red MSX Source (RMS) survey (G326.4755+00.6947; Lumsden et al. 2013). The MMB source associated with the radio emission is mid-infrared dark with a counterpart only appearing at wavelengths of 70 μm and longer. The source itself is coincident with the peak of the strong compact dust emission and so consistent with being a deeply embedded protostellar object. The radio emission is therefore associated with the youngest embedded object, consistent with this radio source being a very young compact H II region. There is no 5 GHz radio counterpart, however it has been covered by the CORNISH-South survey (Irabor et al. 2023) and we can use the mean survey sensitivity of 0.33 mJy (3$\sigma$) to determine a lower limit for the spectral index of +1.14. This source appears to be a very young and embedded object that is associated with an optically thick ionized nebula making it a strong HC H II region candidate.

**G338.566+0.110 (Figure 13b):** The radio source is tightly correlated with the position of the MMB source G338.566+00.110 (Caswell et al. 2011) with an offset of 2.2 arcsec, and coincident with a 70 μm point source, which is itself located towards the peak of the dust emission. There is a 5 GHz counterpart for this radio source detected by the CORNISH-South (G338.5660+0.1096, offset = 0.7 arcsec; Irabor et al. 2023) and, using the integrated flux density (4.98 ± 0.19 mJy) we determine the spectral index to be 0.37 ± 0.03. The coincidence of the methanol-maser emission, far-infrared point source, dust emission and high- and low-frequency radio counterparts all point to this being a strong HC H II region candidate.

14   *A.L Patel et al.*

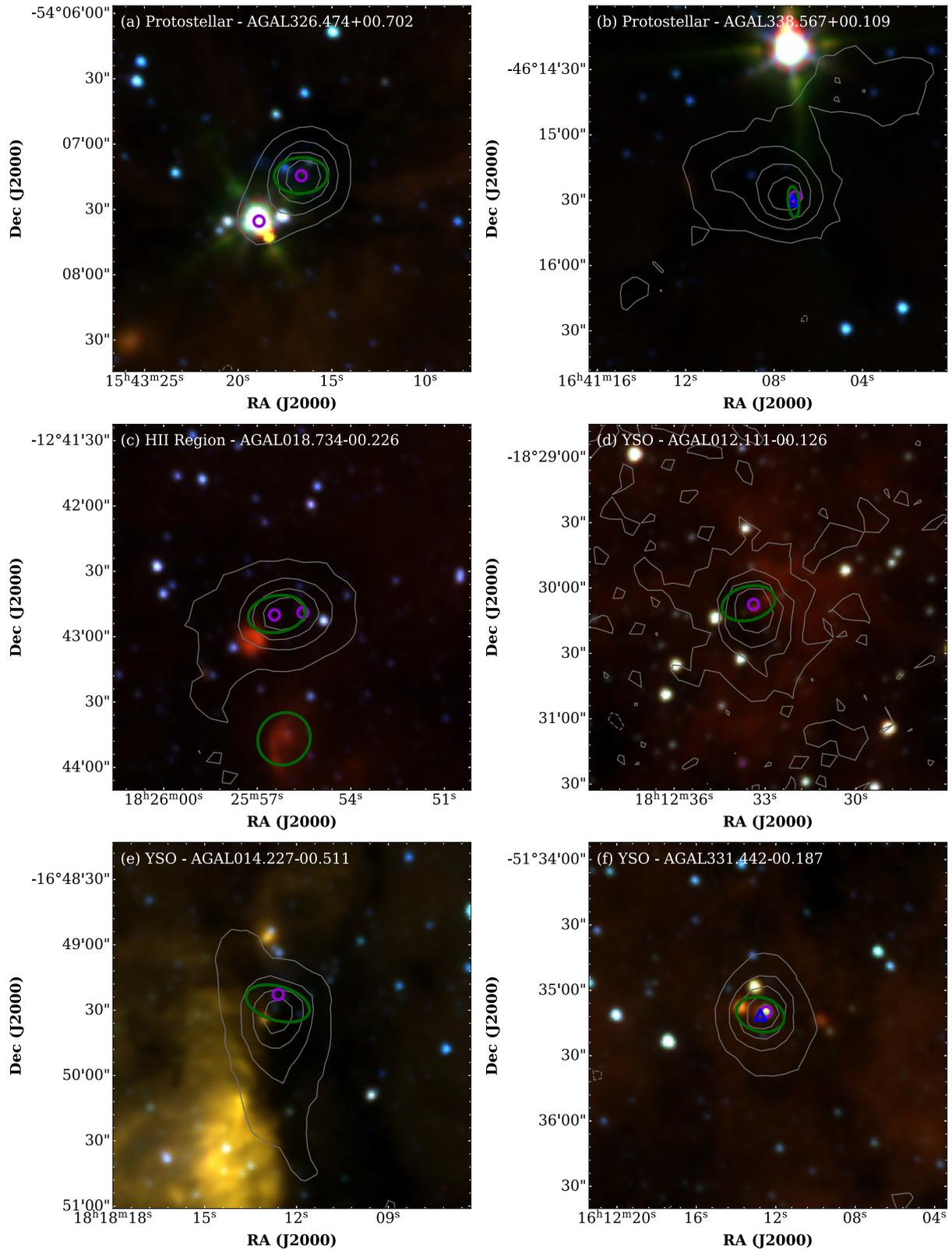

**Figure 13.** Three colour images using IRAC wavebands at 4.5, 5.8 and 8.0 μm for our HC H II region candidates. The green ellipse shows the position of the high-frequency (23.7 GHz) radio source and the purple circles shows the position of the methanol maser. The blue triangles show the position of the low-frequency (5 GHz) radio counterpart and the grey contours trace 870 μm dust emission form ATLASGAL. The ATLASGAL classification is given in the top left.

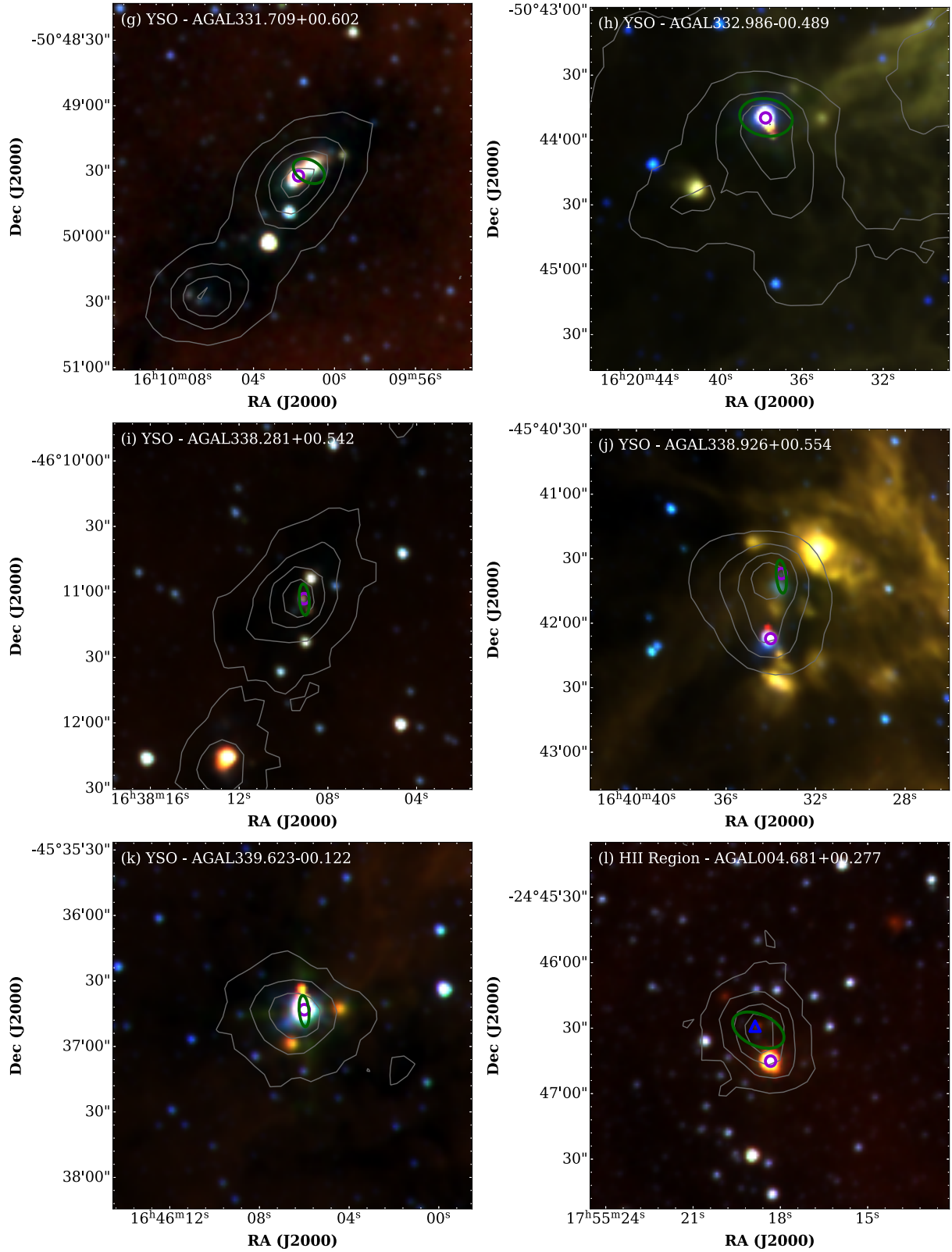

**Figure 13.** Cont.

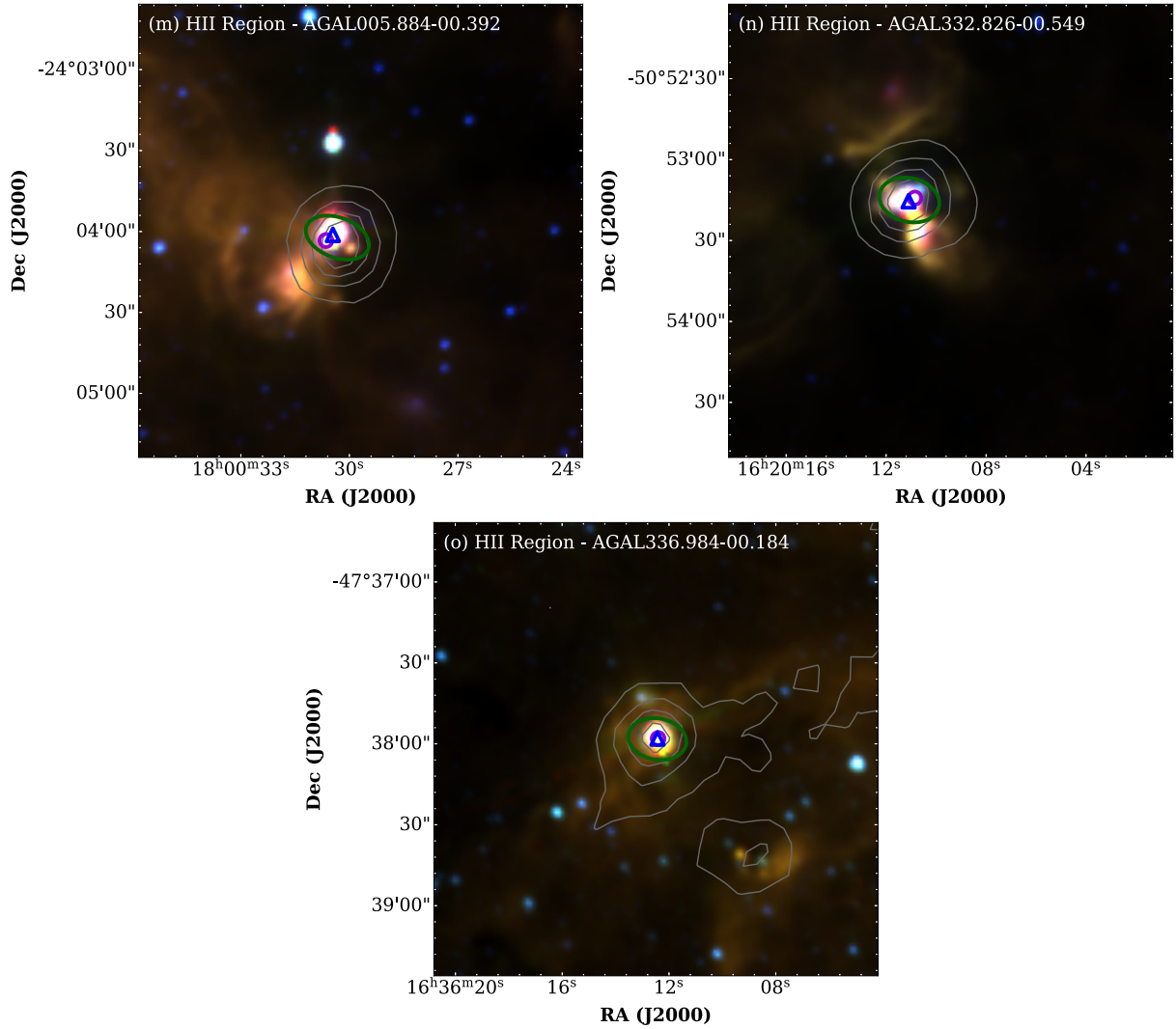

**Figure 13.** Cont.

**G339.681−1.207:** This source has compact high-frequency radio emission that is coincident with two methanol masers G339.681−01.208 and G339.682−01.207 (Caswell et al. 2011) that are offset by 2.4 arcsec and 2.6 arcsec respectively. Both masers are coincident with the peak of the dust emission. This source is located outside the GLIMPSE region and so there are no IRAC data available but there is compact emission at 24 μm (Carey et al. 2009) and 70 μm (Molinari et al. 2010a), which is coincident with the position of the maser and the peak in the dust continuum. This region was observed at 4.8 and 8.6 GHz as part of the RMS survey but no radio emission was detected above 3$\sigma$ at either frequency, i.e. 0.3 mJy and 0.6 mJy respectively (RMS source G339.6816−01.2058; Urquhart et al. 2007) leading the RMS team to classify this source as a YSO. Using the 4.8 GHz radio continuum upper limit we determine a lower limit for the spectral index of +1.16, confirming this object is an excellent HC Hɪɪ region candidate.

**G018.735−0.227 (Figure 13c):** This radio source is one of two within the field in which the radio detection is coincident with the MMB methanol maser G018.735−00.227 (Green et al. 2010) and water maser 18.735−0.227 (Titmarsh et al. 2016). The offset between the radio, methanol, and water maser positions is 1.5 and 1.8 arcsec, respectively and both are coincident with the centre of the compact dust clump. This clump has been classified as an Hɪɪ region by the ATLASGAL team (Urquhart et al. 2022), however, inspection of the 8 μm mid-infrared environment reveals that the extended emission associated with the embedded Hɪɪ region is offset from the high-frequency radio emission, which is located to the south east of the dust clump. No compact IRAC emission is found within the high-frequency radio ellipse, but searching the MIPS-GAL 24 μm catalogue (Gutermuth & Heyer 2015) reveals a discrete point source coincident with the radio emission. Considering that the radio emission is closely associated with both methanol and water maser emission the clump is likely to be associated with outflows and active accretion.

This source was observed at 5 GHz by Hu et al. (2016) who reported the detection of a marginally resolved point source with a peak and integrated flux density of 1.02 mJy beam$^{-1}$ and 1.22 mJy, respectively. This 5 GHz source is located 1.34 arcsec from the maser position and is coincident with the high-frequency radio source. Using the integrated 5 GHz flux density we find the spectral index for this radio source to be 0.20 ± 0.12, indicating this H II region is more likely to be optically thin; however, the lack of a 8 μm source makes this an interesting region and difficult to explain. One possible explanation could be that we are observing a H II region located on the far side of a dense molecular gas cloud. More information is needed to make a reliable determination of the nature of this radio emission.

### 5.2 YSO

**G012.112−0.127 (Figure 13d):** This source is close to the MMB methanol maser G012.112−00.126 (Green et al. 2010) with an offset of only 2.5 arcsec. Strong dust emission from compact clump is coincident with the radio and maser emission. The IRAC images reveal an unresolved mid-infrared point source towards the centre of the clump. This source is not associated with a 5 GHz radio counterpart but we can calculate a lower limit for the spectral index to be −0.87 from an upper limit (0.42 mJy) obtained from the CORNISH data, which is inconclusive. Despite the poor constraint on the spectral index, this source appears to be young, compact and embedded, making it a good HC H II region candidate.

**G014.229−0.510 (Figure 13e):** The detected high-frequency radio emission is close to (offset ~4 arcsec) with the MMB methanol maser G014.230−00.509 (Green et al. 2010) and the peak of the dust emission. The mid-infrared environment is dark at 8 μm and 22 μm but there is a distinct point source at 70 μm, and so this is more likely to be protostellar in origin rather than a YSO as classified by the ATLASGAL team. There is no 5 GHz radio counterpart associated with the high-frequency radio emission, however, the position has been covered by the CORNISH survey and so we are able to determine a lower limit using a upper limit value of 0.37 mJy for the 5 GHz emission. We measure the lower limit to the spectral index to be −0.26, which is inconclusive. The overlap of the star-forming tracers and lack of a mid-infrared counterpart indicate this is a good HC H II region candidate.

**G331.442−0.187 (Figure 13f):** This radio source is close to a mid-infrared point source and MMB methanol maser (offset by 3 arcsec from G331.442−00.187; Caswell et al. 2011). It is positioned near the peak of the dust emission and is associated with 22 μm WISE (Wright et al. 2010) and 70 μm Hi−GaL (Molinari et al. 2010a) emission, confirming that this is an embedded and compact H II region. A 5 GHz counterpart of this radio source is detected in the CORNISH-South data (G331.4420−0.1874, offset = 0.3 arcsec; Irabor et al. 2023. Using the integrated flux density, we determine the spectral index to be 0.08 ± 0.06, which indicates the nebula is optically thin and that this is more likely to be an UC H II region.

**G331.710+0.604 (Figure 13g):** This source is offset (5.3 arcsec) from the MMB methanol maser G331.710+00.603 (Caswell et al. 2011) and coincident with the peak of the dust emission and a 8 μm mid-infrared point source. It has no 5 GHz radio counterpart, however it has been covered by the CORNISH-South survey (Irabor et al. 2023) and we can use the mean survey sensitivity of 0.33 mJy ($3\sigma$) to determine a lower limit for the spectral index of +0.64. The coincidence of dust and embedded mid-infrared sources is consistent with this being a deeply embedded young star also being associated with high-frequency radio emission makes this source a good HC H II region candidate.

**G332.987−0.487 (Figure 13h):** The high-frequency radio emission is tightly correlated with a methanol maser G332.987−00.487 (offset = 1.8 arcsec; Caswell et al. 2011) and a mid-infrared point source, but is slightly offset from the peak of the dust emission. Large-scale extended emission is found to the north and west of the radio and MMB positions. These is no 5 GHz counterpart for this source, however, we can calculate a lower limit of +1.74 from the CORNISH-South survey sensitivity ($3\sigma$ = 0.33 mJy; Irabor et al. 2023), consistent with the radio emission being associated with an optically thick H II region. This is therefore is an excellent HC H II region candidate.

**G338.280+0.542 (Figure 13i):** This high-frequency radio source is tightly associated with MMB maser G338.280+00.542 (offset < 1 arcsec; Caswell et al. 2011) and coincident with the peak of the dust emission. The radio emission is mid-infrared dark at 8 μm, however, at 22 μm we see two distinct point sources. There is no 5 GHz radio counterpart, however, there are upper limits available at 4.8 and 8.6 GHz from the RMS survey (i.e. $3\sigma$ upper limits are 0.4 mJy and 1.2 mJy respectively; Urquhart et al. 2007) from which we can determine a value of +1.57 as a lower limit to the spectral index. The coincidence of this radio source with the star-formation tracers and the lower limit for the spectral index make this a strong HC H II region candidate.

**G338.925+0.556 (Figure 13j):** The mid-infrared image reveals a larger extended structure to the west. The radio source is coincident with a dense dust clump located on the edge of the extended mid-infrared source and the methanol maser G338.925+00.557; (Caswell et al. 2011), which is offset by 1.0 arcsec. Inspection of the mid-infrared image reveals a small group of point sources located towards the centre of the clump, indicating the presence of a small protocluster. There is no 5 GHz counterpart but, again using archival data from the RMS survey, we estimate the $3\sigma$ upper limits of 0.5 mJy and 0.5 mJy at 4.8 and 8.6 GHz, respectively (RMS source G338.9237+00.5618; Urquhart et al. 2007), and use these to determine a lower limit of +2.29 for the spectral index. This source is therefore an excellent HC H II region candidate.

**G339.622−0.121 (Figure 13k):** The high-frequency radio emission is coincident with a mid-infrared bright point source, which itself is located towards the centre of the dust clump and is associated with the MMB methanol maser G339.622−00.121 (Caswell et al. 2011) that is offset by 0.7 arcsec. There is no 5 GHz counterpart for this source. However, again using the CORNISH-South survey sensitivity ($3\sigma$=0.33 mJy; Irabor et al. 2023), we calculate a lower limit for the spectral index of +1.42. The spectral index and coincidence of the radio emission, methanol maser, dense gas and mid-infrared point sources all make this an excellent HC H II region candidate.

### 5.3 Optically thick H II regions

Below we discuss four sources identified in this paper and by the ATLASGAL team as being H II regions that have a positive spectral index, making them potential HC H II region candidates.

**G004.680+0.277 (Figure 13l):** The radio source is located towards the centre of ATLASGAL clump AGAL004.681+00.277 (Contreras et al. 2013) but is offset by 15.3 arcsec to the north from the central maser position. This is tightly correlated with a

18    *A.L Patel et al.*

**Table 7.** Summary of the physical properties for all the ATLASGAL clumps hosting HC H II region candidates.

| Radio name | CSC name | Log ($L_{bol}$) ($L_\odot$) | Log ($M_{fwhm}$) ($M_\odot$) | $L_{bol}/M_{fwhm}$ ($L_\odot/M_\odot$) | ATLASGAL Classification |
|---|---|---|---|---|---|
| G004.680+0.277 | AGAL004.681+00.277 | 3.35 | 1.95 | 25.2 | H II region |
| G005.885−0.392 | AGAL005.884−00.392 | 5.33 | 2.72 | 403 | H II region |
| G012.112−0.127 | AGAL012.111−00.126 | 3.18 | 2.24 | 8.58 | YSO |
| G014.229−0.510 | AGAL014.227−00.511 | 3.18 | 2.18 | 10.1 | YSO |
| G018.735−0.227 | AGAL018.734−00.226 | 4.81 | 3.79 | 10.3 | H II region |
| G326.475+0.703 | AGAL326.474+00.702 | 3.73 | 2.41 | 20.9 | Protostellar |
| G331.442−0.187 | AGAL331.442−00.187 | 3.91 | 2.64 | 18.5 | YSO |
| G331.710+0.604 | AGAL331.709+00.602 | 4.74 | 3.71 | 10.9 | YSO |
| G332.826−0.549 | AGAL332.826−00.549 | 5.54 | 3.09 | 283 | H II region |
| G332.987−0.487 | AGAL332.986−00.489 | | | | YSO |
| G336.983−0.183 | AGAL336.984−00.184 | 4.24 | 2.06 | 154 | H II region |
| G338.280+0.542 | AGAL338.281+00.542 | 3.62 | 2.59 | 10.6 | YSO |
| G338.566+0.110 | AGAL338.567+00.109 | 3.74 | 2.72 | 10.5 | Protostellar |
| G338.925+0.556 | AGAL338.926+00.554 | 4.96 | 3.51 | 28.5 | YSO |
| G339.622−0.121 | AGAL339.623−00.122 | 3.87 | 2.28 | 38.2 | YSO |
| G339.681−1.207 | AGAL339.681−01.209 | 4.07 | 2.31 | 56.9 | Protostellar |

bright mid-infrared point source. Additionally, there is a weak mid-infrared point source located towards the centre of the dust clump, that is coincident with the position of the high frequency radio ellipse. We have looked at the corresponding 24 μm MIPSGAL data (Carey et al. 2009; Gutermuth & Heyer 2015), which also reveals the presence of a discrete point source towards the centre of the dust clump and is coincident with the IRAC and radio sources. The centre of the clump is also host to a 5 GHz radio counterpart allowing us to determine a spectral index of +0.39 ± 0.07, which is marginally optically thick. Despite this source being below the estimated optical depth threshold the correspondence of compact dust emission, the weak mid-infrared emission and positive spectral index suggests it is in an early stage of its evolution and for these reasons we consider this source to be a HC H II region candidate.

**G005.885−0.392 (Figure 13m):** This 23.7 GHz radio detection is part of a more extended, filamentary type structure and is a well known UC H II region (Bronfman et al. 1996). The radio emission overlaps with MMB methanol maser G005.885−00.393 (offset by 4.5 arcsec; Caswell 2010) and water maser 5.885−0.393 (offset by 2.4 arcsec; Titmarsh et al. 2016). The high-frequency radio source has a 5 GHz radio counterpart (5.885−0.392; Becker et al. 1994) that is offset by 1.9 arcsec and is coincident with the peak of the dust emission. We calculate the spectral index to be +0.81 ± 0.02 indicating the emission is optically thick at 23.7 GHz. Inspection of the mid-infrared environment for this dust clump reveals that the high-frequency radio detection is embedded within a complicated and somewhat extended H II region. Despite this, with the data we have available, the 23.7 GHz radio emission is still unresolved and positionally correlated to all the star forming tracers discussed and is therefore a highly interesting source for further investigation.

**G332.826−0.549 (Figure 13n):** The high-frequency radio emission is coincident with the peak of the strong compact dust emission and is correlated with a MMB methanol maser G332.826−00.549 (Caswell et al. 2011); offset 2.2 arcsec. The centre of the clump is also host to a 5 GHz radio counterpart G332.8260−00.5493 (Urquhart et al. 2007 offset; 0.5 arcsec) allowing us to determine a spectral index of +0.77 ± 0.07. The mid-infrared image reveals a complicated structure but the radio source and maser are coincident with a bright 8 μm point source. Based on the asso-ciations explored we consider this a strong HC H II region candidate.

**G336.983−0.183 (Figure 13o):** The high frequency radio emission is located towards the centre of the dust clump and is associated with MMB methanol maser G336.983−00.183 (Caswell et al. 2011) and a 5 GHz radio source G336.9833−00.1832 (Urquhart et al. 2007) both offset by 1.1 arcsec. Inspection of the mid-infrared image reveals that the radio emission is coincident with a mid-infrared point source at 8 μm. We measure the spectral index of the emission to be +0.72 ± 0.03, confirming the ionized nebula is optically thick. From the close correlation of star formation tracers and spectral index measurement, we can say this source is a strong HC H II region candidate.

After reviewing all the available evidence we find 13 sources that fit the criteria for being HC H II regions. We will follow up these sources at a higher resolution to determine their physical properties and their true nature.

## 6 SUMMARY AND CONCLUSIONS

We report the results of an analysis of high-frequency radio continuum archival data towards a sample of methanol masers identified by the MMB survey (Caswell 2010; Green et al. 2010; Caswell et al. 2011) and located between Galactic longitudes 310° and 20°. The original observations were made using the ATCA and provide continuum coverage from 21.3 to 24.7 GHz. These have produced high-sensitivity images (typical rms noise value of ~ 0.2 mJy) centred at 23.7 GHz. The main aim of this work is to identify new HC H II region candidates and is based on the hypothesis that methanol masers are an excellent tracer of young and embedded high-mass stars.

We have imaged 128 distinct fields covering the positions of 141 methanol masers from which we have been able to identify 68 discrete radio sources. We have conducted a multi-wavelength analysis to determine the nature of these radio detections including mid-infrared images from GLIMPSE (Benjamin et al. 2003), WISE (Wright et al. 2010) and MIPSGAL (Carey et al. 2009; Gutermuth & Heyer 2015), far-infrared and submillimetre from HiGAL (Molinari et al. 2010a) and ATLASGAL (Schuller et al. 2009) and radio continuum from the RMS survey (Lumsden et al. 2013; Urquhart et al. 2007) and CORNISH surveys (Hoare et al.

2012; Purcell et al. 2013; Irabor et al. 2023). This has resulted in the identification of 49 H II regions.

Our main findings are as follows:

- We find that 23 per cent of the observed MMB methanol masers in this study are associated with high frequency radio emission within 6 arcsec. A similar studies conducted by Urquhart et al. 2013a,b and Nguyen et al. (2022) report association statistics of between 12 and 16 per cent for MMB methanol masers and low-frequency radio emission (∼5 GHz). This leads us to conclude that we are detecting a significant number of optically thick H II regions not previously detected in 5 GHz surveys.

- The majority of the H II region identified (76 per cent) have been previously identified as such in the literature, however, the remaining H II regions were previously classified as YSOs or protostellar. Analysis of the spectral index for many of these deeply embedded objects reveals them to be optically thick between 5 and 23.7 GHz, making these excellent HC H II candidates.

- We have investigated the correlation between the luminosity-to-mass ratio and luminosity and find that methanol maser sites that have a luminosity > $10^3 \, \mathrm{L_\odot}$ and a $L_{\mathrm{bol}}/M_{\mathrm{clump}}$-ratio of > ∼ 5.1 $\mathrm{L_\odot \, M_\odot^{-1}}$ host H II regions. The embedded objects must be luminous enough to produce an H II region. The minimum bolometric luminosity is consistent with that of a early B-type zero age main sequence star (∼ $10^3 \, \mathrm{L_\odot}$), thus supporting previous studies that have reported that methanol masers are exclusively associated with high-mass star formation. These minimum thresholds can be used to better select MMB samples to increase the detection rate for high-frequency radio emission in future observations.

- Of the 49 H II regions identified in this study we find 13 strong HC H II region candidates based on their mid-infrared environments and spectral index measurements.

　　This is the first in a series of four papers to identify the role HC H II regions play in the massive-star formation sequence. Overall, this study has been successful in identifying H II regions and found a significant number of very good HC H II region candidates by looking for high frequency radio emission towards MMB methanol masers. These will be followed up at a higher-resolution to confirm their nature and determine their physical properties.

## ACKNOWLEDGEMENTS

We thank the reviewer for taking the time to provide detailed and concise comments. We appreciate all the valuable comments and suggestions, that have helped us in improving the quality of this manuscript. A. L. Patel wishes to acknowledge an STFC (Science and Technology Facilities Council) PhD studentship for this work. ALP further acknowledges the support from others in proofreading and providing comments to the early drafts of the manuscript. This research made use of `ASTROPY`[6] a community-developed core `PYTHON` package for Astronomy (Astropy Collaboration et al. 2013, 2018); `MATPLOTLIB` (Hunter 2007). This document was prepared using the Overleaf web application, which can be found at www.overleaf.com.

---

[6] https://www.astropy.org

## DATA AVAILABILITY

The data underlying this article are available in the article and in its online supplementary material.

## References

Anglada G., Estalella R., Pastor J., Rodriguez L. F., Haschick A. D., 1996, ApJ, 463, 205
Astropy Collaboration et al., 2013, A&A, 558, A33
Astropy Collaboration et al., 2018, AJ, 156, 123
Becker R. H., White R. L., Helfand D. J., Zoonematkermani S., 1994, ApJS, 91, 347
Beltrán M. T., Cesaroni R., Moscadelli L., Codella C., 2007, A&A, 471, L13
Benjamin R. A., et al., 2003, PASP, 115, 953
Beuther H., Walsh A., Schilke P., Sridharan T. K., Menten K. M., Wyrowski F., 2002, A&A, 390, 289
Beuther H., et al., 2016, A&A, 595, A32
Bihr S., et al., 2016, A&A, 588, A97
Billington S. J., et al., 2019, MNRAS, 490, 2779
Bonnell I. A., Clarke C. J., Bate M. R., Pringle J. E., 2001, MNRAS, 324, 573
Breen S. L., Ellingsen S. P., Contreras Y., Green J. A., Caswell J. L., Stevens J. B., Dawson J. R., Voronkov M. A., 2013, MNRAS, 435, 524
Breen S. L., et al., 2015, MNRAS, 450, 4109
Bronfman L., Nyman L.-A., May J., 1996, A&AS, 115, 81
Brunthaler A., et al., 2021, A&A, 651, A85
Carey S. J., et al., 2009, PASP, 121, 76
Carpenter J. M., Snell R. L., Schloerb F. P., 1990, ApJ, 362, 147
Caswell J. L. e., 2010, MNRAS, 404, 1029
Caswell J. L., et al., 2011, MNRAS, 417, 1964
Cesaroni R., Sánchez-Monge Á., Beltrán M. T., Molinari S., Olmi L., Treviño-Morales S. P., 2016, A&A, 588, L5
Churchwell E., 2002, ARA&A, 40, 27
Comeron F., Torra J., 1996, A&A, 314, 776
Contreras Y., et al., 2013, A&A, 549, A45
Cragg D. M., Johns K. P., Godfrey P. D., Brown R. D., 1992, MNRAS, 259, 203
Davies B., Hoare M. G., Lumsden S. L., Hosokawa T., Oudmaijer R. D., Urquhart J. S., Mottram J. C., Stead J., 2011, MNRAS, 416, 972
Deacon R. M., Chapman J. M., Green A. J., Sevenster M. N., 2007, ApJ, 658, 1096
Dyson J. E., Williams R. J. R., Redman M. P., 1995, Monthly Notices of the Royal Astronomical Society, 277, 700
Dzib S. A., et al., 2023, A&A, 670, A9
Evans II N. J., 1999, ARA&A, 37, 311
Fazio G. G., et al., 2004, ApJS, 154, 10
Gaume R. A., Goss W. M., Dickel H. R., Wilson T. L., Johnston K. J., 1995, ApJ, 438, 776
Genzel R., Harris A. I., Stutzki J., 1989, in Böhm-Vitense E., ed., Infrared Spectroscopy in Astronomy. p. 115
González-Avilés M., Lizano S., Raga A. C., 2005, ApJ, 621, 359
Green J. A., et al., 2009, MNRAS, 392, 783
Green J. A., et al., 2010, MNRAS, 409, 913
Green J. A., et al., 2012, MNRAS, 420, 3108
Gutermuth R. A., Heyer M., 2015, AJ, 149, 64
Hoare M. G., Kurtz S. E., Lizano S., Keto E., Hofner P., 2007, Protostars and Planets V, pp 181–196
Hoare M. G., et al., 2012, PASP, 124, 939
Hosokawa T., Yorke H. W., Omukai K., 2010, ApJ, 721, 478
Hu B., Menten K. M., Wu Y., Bartkiewicz A., Rygl K., Reid M. J., Urquhart J. S., Zheng X., 2016, ApJ, 833, 18
Hunter J. D., 2007, Computing in Science & Engineering, 9, 90
Irabor T., et al., 2023, MNRAS, 520, 1073
Jones B. M., et al., 2020, MNRAS, 493, 2015
Keto E., Klaassen P., 2008, The Astrophysical Journal, 678, L109–L112

Kurtz S., 2005, in Lis D. C., Blake G. A., Herbst E., eds, IAU Symposium Vol. 231, Astrochemistry: Recent Successes and Current Challenges. pp 47–56, doi:10.1017/S1743921306007034
Kurtz S., Hofner P., 2005, AJ, 130, 711
Lumsden S. L., Hoare M. G., Urquhart J. S., Oudmaijer R. D., Davies B., Mottram J. C., Cooper H. D. B., Moore T. J. T., 2013, ApJS, 208, 11
Martins F., Schaerer D., Hillier D. J., 2005, A&A, 436, 1049
McKee C. F., Ostriker E. C., 2007, ARA&A, 45, 565
McKee C. F., Tan J. C., 2002, Nature, 416, 59
McKee C. F., Tan J. C., 2003, ApJ, 585, 850
Medina S. N. X., et al., 2019, A&A, 627, A175
Menten K. M., Reid M. J., Pratap P., Moran J. M., Wilson T. L., 1992, ApJ, 401, L39
Minier V., Ellingsen S. P., Norris R. P., Booth R. S., 2003, A&A, 403, 1095
Molinari S., Brand J., Cesaroni R., Palla F., 1996, A&A, 308, 573
Molinari S., et al., 2010a, PASP, 122, 314
Molinari S., et al., 2010b, A&A, 518, L100
Motte F., Bontemps S., Louvet F., 2018, ARA&A, 56, 41
Nguyen H., et al., 2022, A&A, 666, A59
Panagia N., 1973, AJ, 78, 929
Purcell C. R., et al., 2013, ApJS, 205, 1
Purser S. J. D., et al., 2016, MNRAS, 460, 1039
Purser S. J. D., Lumsden S. L., Hoare M. G., Kurtz S., 2021, MNRAS, 504, 338
Sánchez-Monge Á., Beltrán M. T., Cesaroni R., Fontani F., Brand J., Molinari S., Testi L., Burton M., 2013, A&A, 550, A21
Sault R. J., Teuben P. J., Wright M. C. H., 1995, in Shaw R. A., Payne H. E., Hayes J. J. E., eds, Astronomical Society of the Pacific Conference Series Vol. 77, Astronomical Data Analysis Software and Systems IV. p. 433 (arXiv:astro-ph/0612759)
Schuller F., et al., 2009, A&A, 504, 415
Thompson R. I., 1984, ApJ, 283, 165
Thompson M. A., Urquhart J. S., Moore T. J. T., Morgan L. K., 2012, MNRAS, 421, 408
Titmarsh A. M., Ellingsen S. P., Breen S. L., Caswell J. L., Voronkov M. A., 2014, MNRAS, 443, 2923
Titmarsh A. M., Ellingsen S. P., Breen S. L., Caswell J. L., Voronkov M. A., 2016, MNRAS, 459, 157
Urquhart J. S., Busfield A. L., Hoare M. G., Lumsden S. L., Clarke A. J., Moore T. J. T., Mottram J. C., Oudmaijer R. D., 2007, A&A, 461, 11
Urquhart J. S., Morgan L. K., Thompson M. A., 2009, A&A, 497, 789
Urquhart J. S., et al., 2013a, MNRAS, 431, 1752
Urquhart J. S., et al., 2013b, MNRAS, 435, 400
Urquhart J. S., et al., 2014, MNRAS, 443, 1555
Urquhart J. S., et al., 2018, MNRAS, 473, 1059
Urquhart J. S., et al., 2022, MNRAS, 510, 3389
Walsh A. J., Burton M. G., Hyland A. R., Robinson G., 1998, MNRAS, 301, 640
Walsh A. J., et al., 2011, MNRAS, 416, 1764
Whitworth A. P., Bhattal A. S., Chapman S. J., Disney M. J., Turner J. A., 1994, MNRAS, 268, 291
Wilson W. E., et al., 2011, MNRAS, 416, 832
Wood D. O. S., Churchwell E., 1989, ApJS, 69, 831
Wright E. L., et al., 2010, AJ, 140, 1868
Yang A. Y., Thompson M. A., Tian W. W., Bihr S., Beuther H., Hindson L., 2019, MNRAS, 482, 2681
Yang A. Y., et al., 2021, Astronomy & Astrophysics, 645, A110
Yorke H. W., Bodenheimer P., 1999, ApJ, 525, 330
Zinnecker H., Yorke H. W., 2007, ARA&A, 45, 481
de Villiers H. M., et al., 2015, MNRAS, 449, 119

## APPENDIX A: 23.7 GHZ RADIO MAPS AND THREE COLOUR IMAGES

In Fig. A1 we present the 23.7 GHz radio maps and three colour images for all the detected radio sources in this study.

## APPENDIX B: 5 GHZ RADIO MAPS

In Fig. B1 we show the 5 GHz radio maps for the corresponding MAGPIS counterparts. Fig. B2 shows the 5 GHz radio maps for the corresponding CORNISH North and South counterparts. Fig. B3 shows the 5 GHz radio maps for the corresponding RMS counterparts and in Fig. B4 we present the 5 GHz radio maps for the nine non-detections.

This paper has been typeset from a TeX/LaTeX file prepared by the author.

22   *A.L Patel et al.*

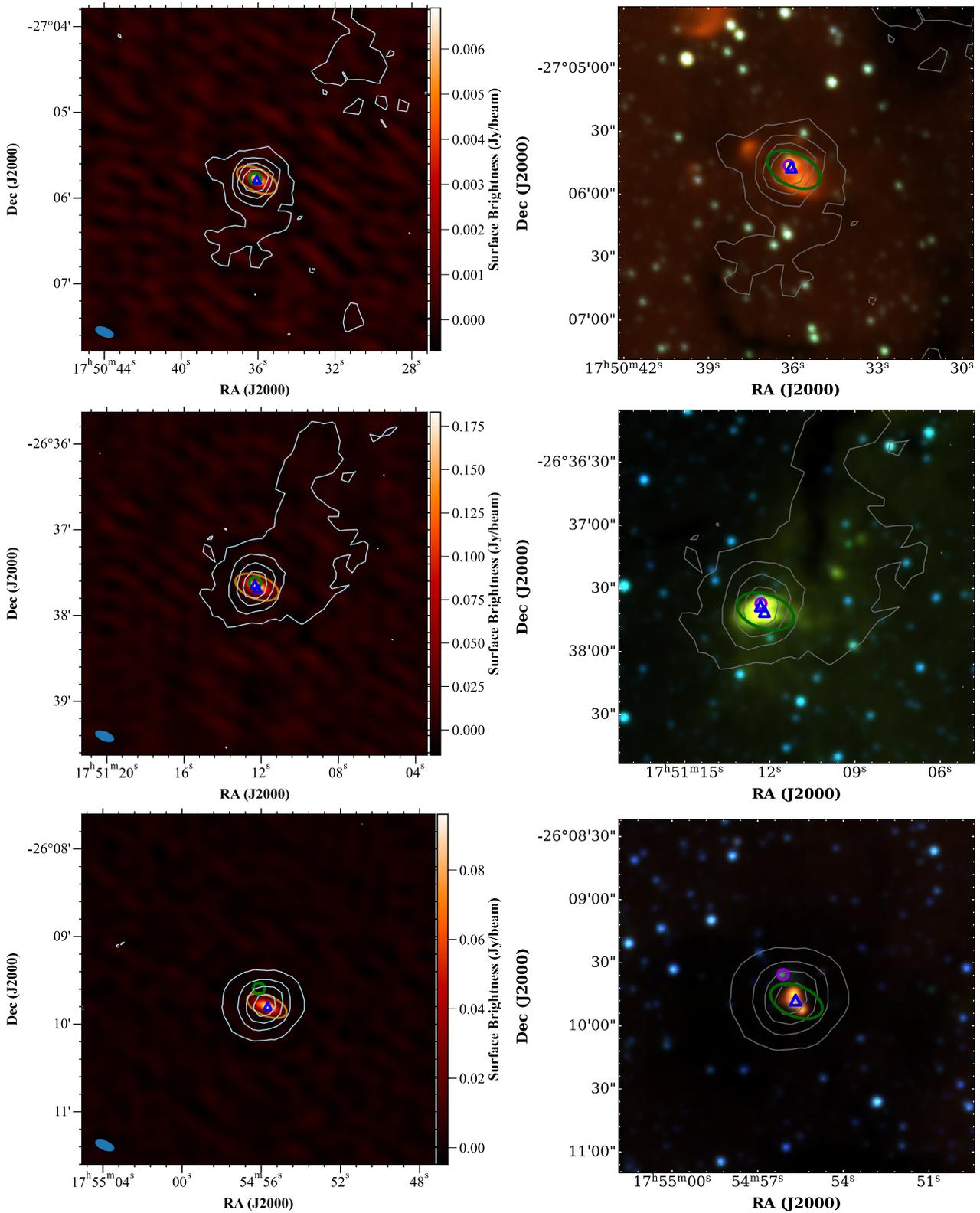

**Figure A1.** Left panel: High frequency radio maps of the detected sources above $3\sigma$. The orange ellipse shows the resultant fit of the radio emission. Green circles show the position(s) of the methanol masers in the field. Blue triangles represent the position of 5 GHz radio counterparts. The grey contours trace the 870 μm dust emission from ATLASGAL. The filled blue ellipse in the bottom left hand corner of each image indicates the size and orientation of the synthesised beam. Right panel: Three colour images of the corresponding radio sources using the 4.5, 5.8 & 8.0 μm IRAC bands from the GLIMPSE survey. The green ellipse shows the position of the high-frequency (23.7 GHz) radio source and the purple circles shows the position of the methanol maser. The blue triangles show the position of the low-frequency (5 GHz) radio counterpart, the grey contours trace 870 μm dust emission from ATLASGAL (Schuller et al. 2009). The radio and clump name for each source is given in the caption. *Top row*: G002.143+0.009, AGAL002.142+00.009, *Middle row*: G002.614+0.134, AGAL002.616+00.134, *Bottom row*: G003.438−0.349, AGAL003.439−00.349.

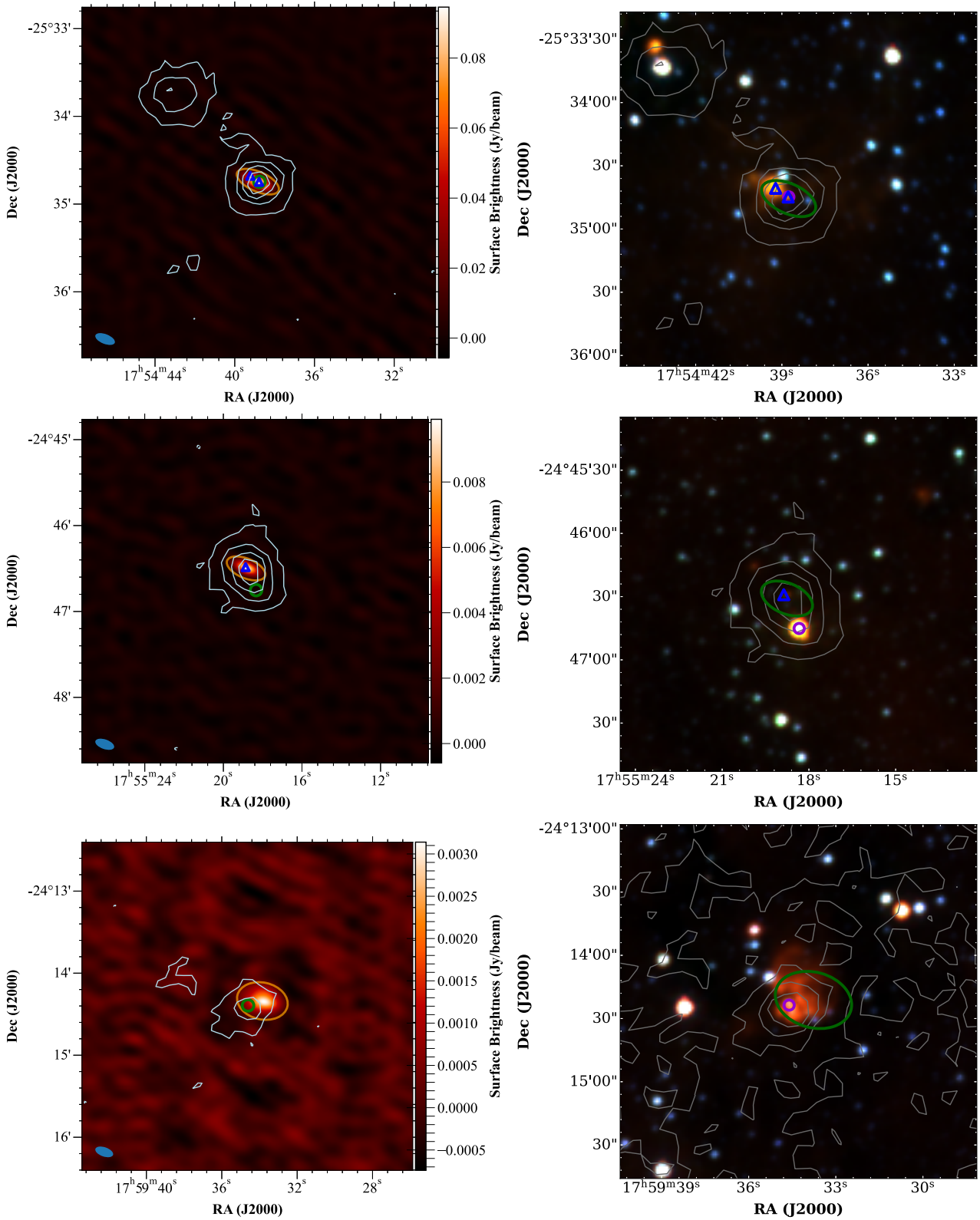

**Figure A1.** Cont. *Top row*: G003.910+0.001, AGAL003.911+00.001, *Middle row*: G004.680+0.277, AGAL004.681+00.277, *Bottom row*: G005.629−0.291, AGAL005.629−00.294.

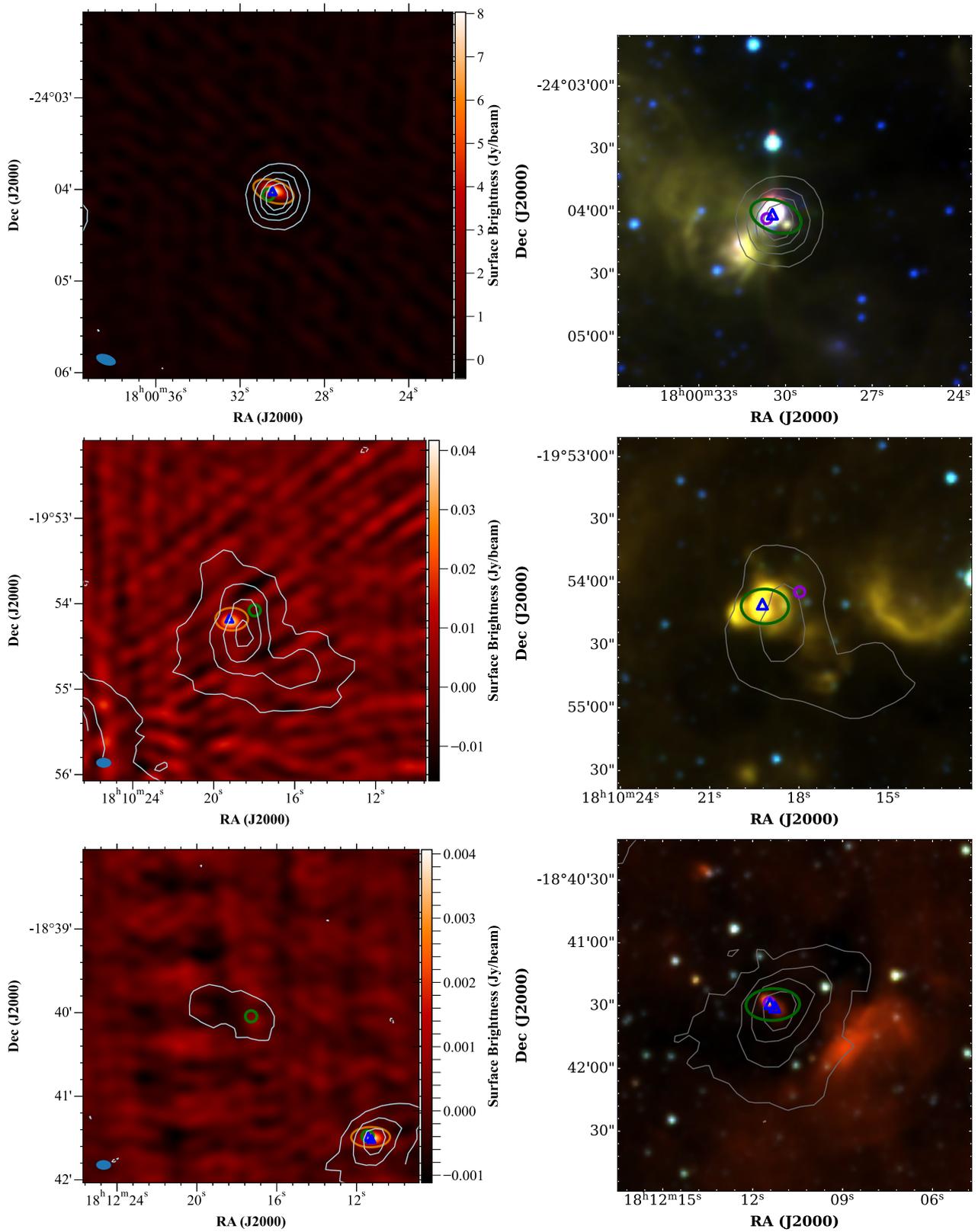

**Figure A1.** Cont. *Top row*: G005.885−0.392, AGAL005.884−00.392, *Middle row*: G010.629−0.338, AGAL010.626−00.337, *Bottom row*: G011.904−0.141, AGAL011.902−00.141.

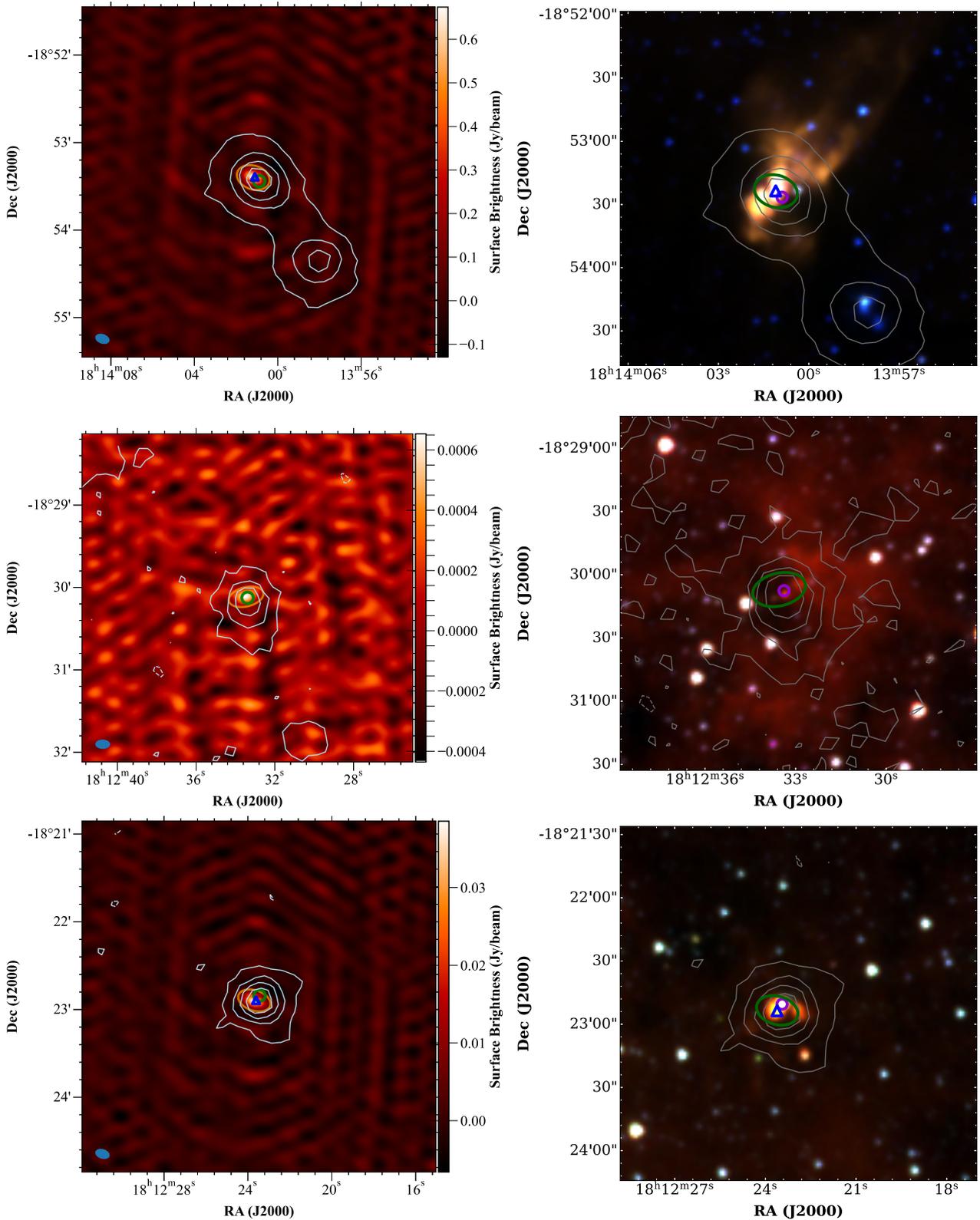

**Figure A1.** Cont. *Top row*: G011.937−0.616, AGAL011.936−00.616, *Middle row*: G012.112−0.127, AGAL012.111−00.126, *Bottom row*: G012.199−0.034, AGAL012.198−00.034.

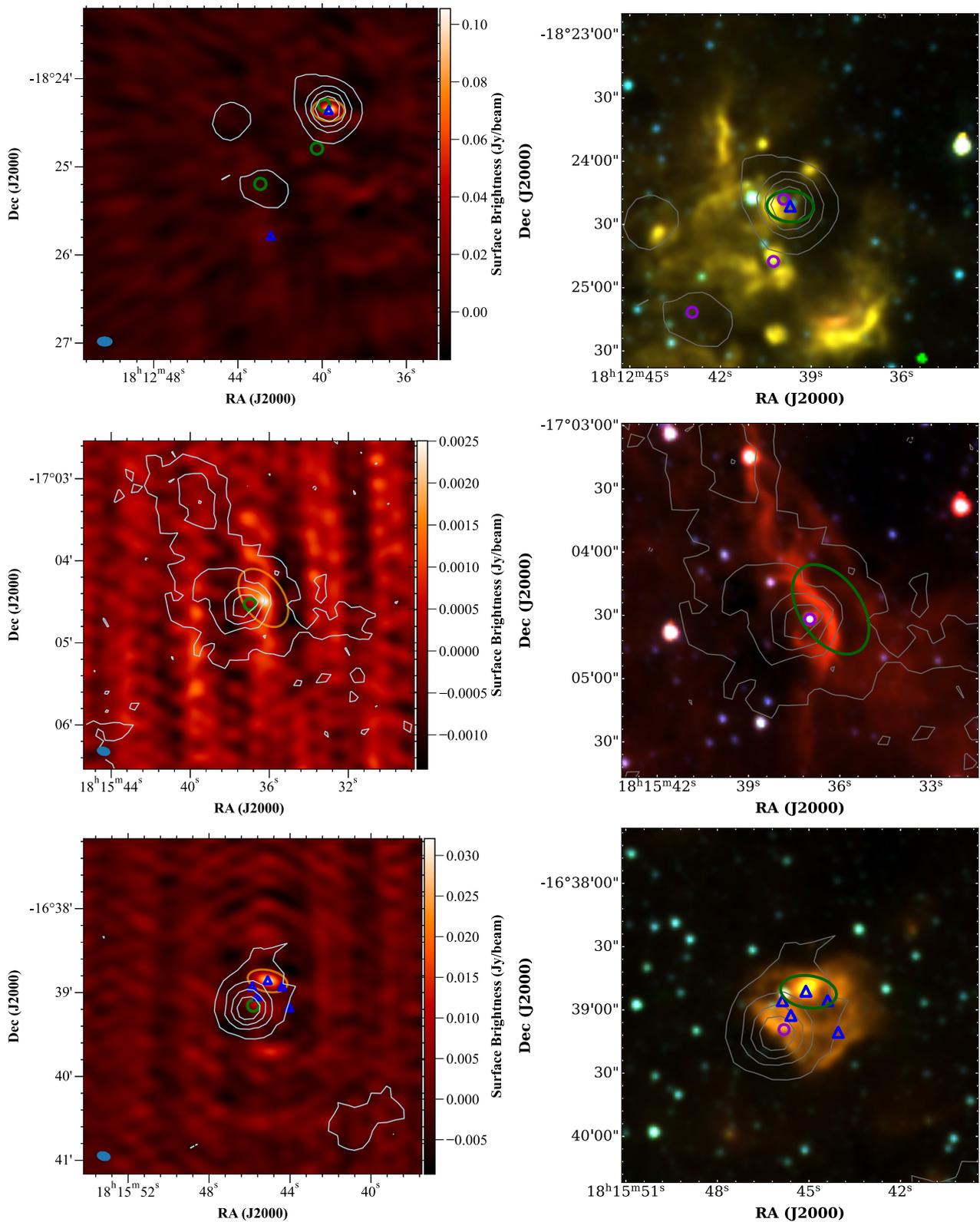

**Figure A1.** Cont. *Top row*: G012.208−0.102, AGAL012.208−00.102, *Middle row*: G013.713−0.080, AGAL013.714−00.084, *Bottom row*: G014.104+0.092, AGAL014.101+00.086.

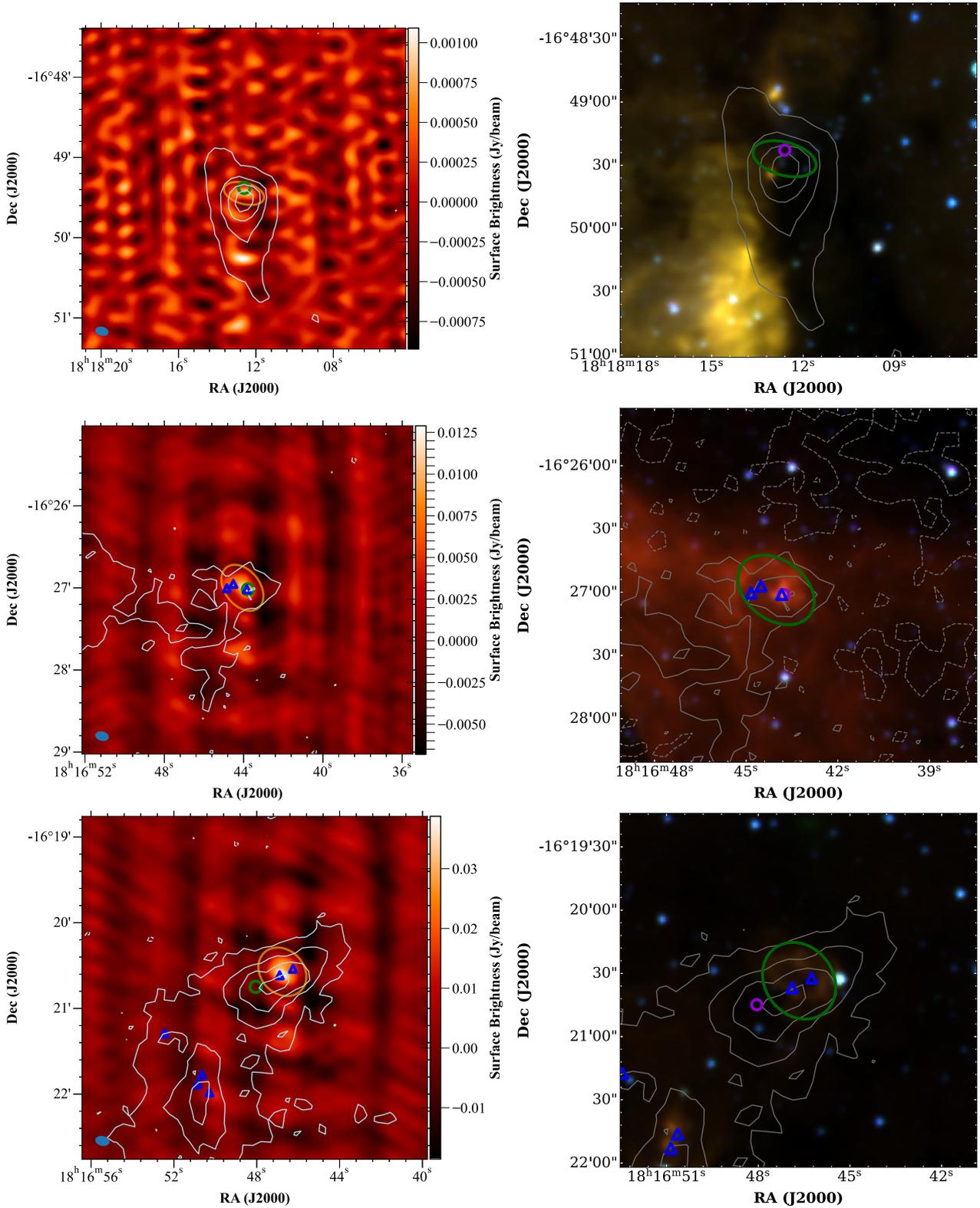

**Figure A1.** Cont. *Top row*: G014.229−0.510, AGAL014.227−00.511, *Middle row*: G014.391−0.021, AGAL014.389−00.019, *Bottom row*: G014.490+0.021, AGAL014.489+00.017.

28   *A.L Patel et al.*

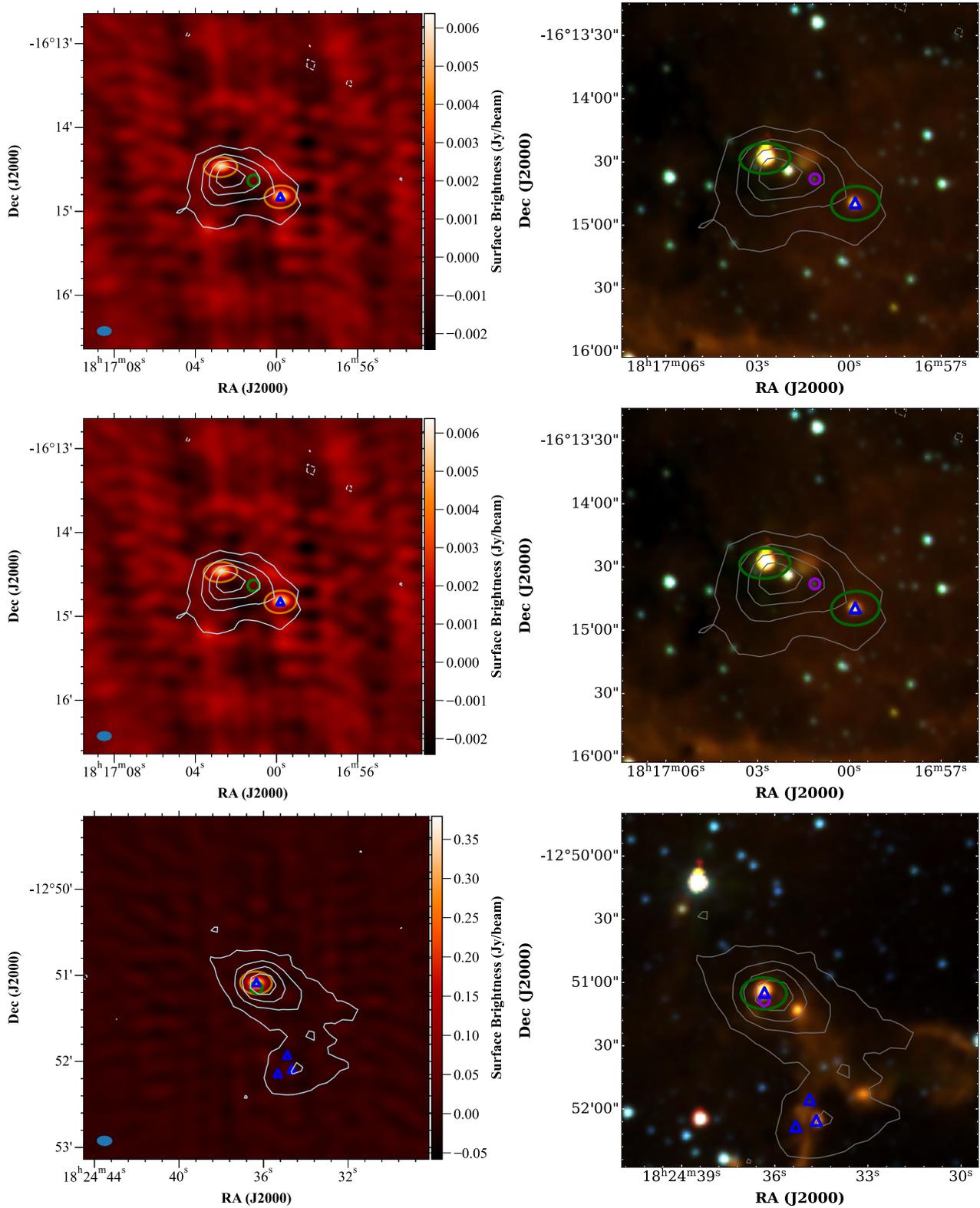

**Figure A1.** Cont. *Top row*: G014.599+0.020, AGAL014.607+00.012, *Middle row*: G014.610+0.012, AGAL014.607+00.012, *Bottom row*: G018.461−0.004, AGAL018.461−00.002.

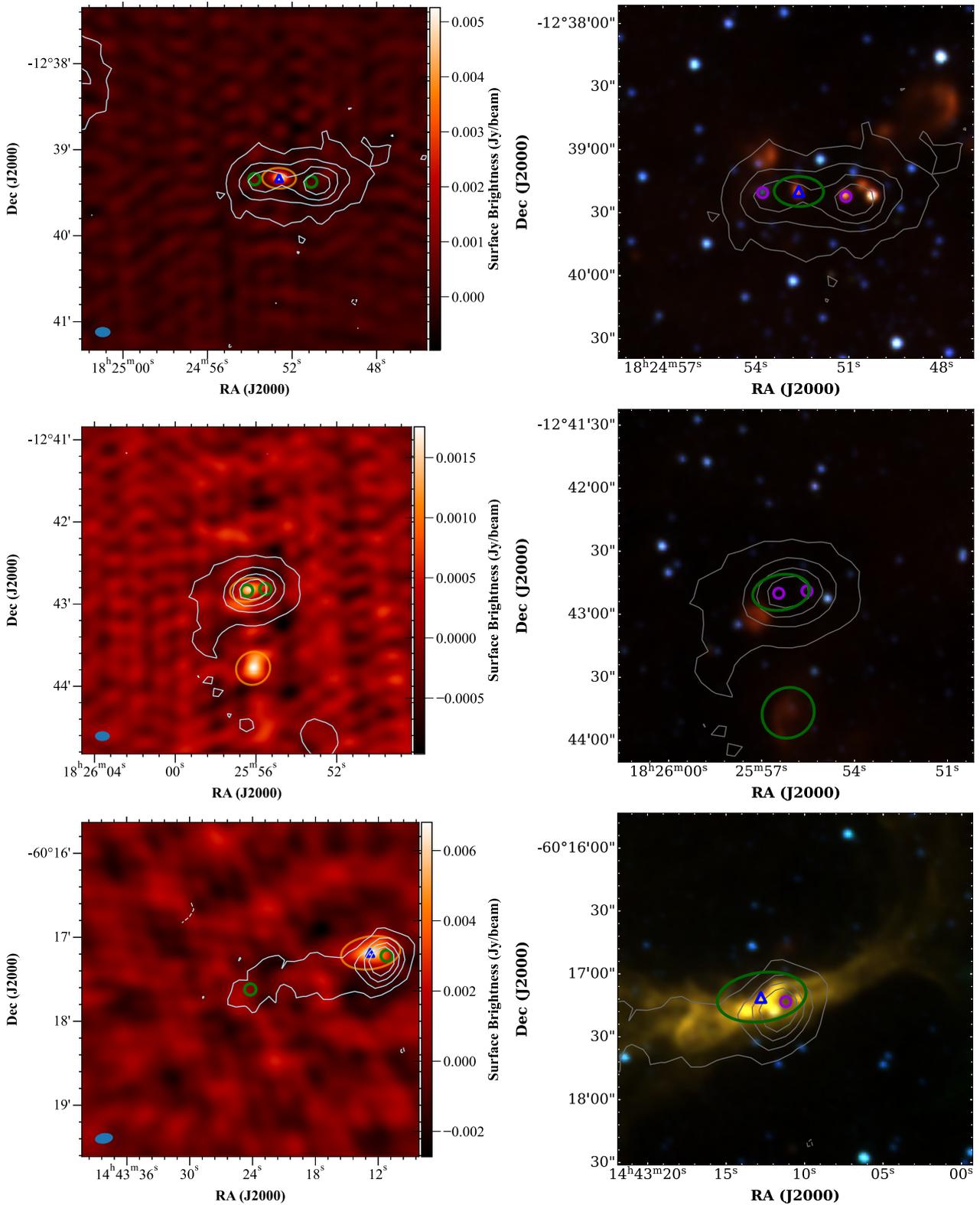

**Figure A1.** Cont. *Top row*: G018.665+0.029, AGAL018.666+00.026, *Middle row*: G018.735−0.227, AGAL018.734−00.226, *Bottom row*: G316.363−0.362, AGAL316.361−00.362.

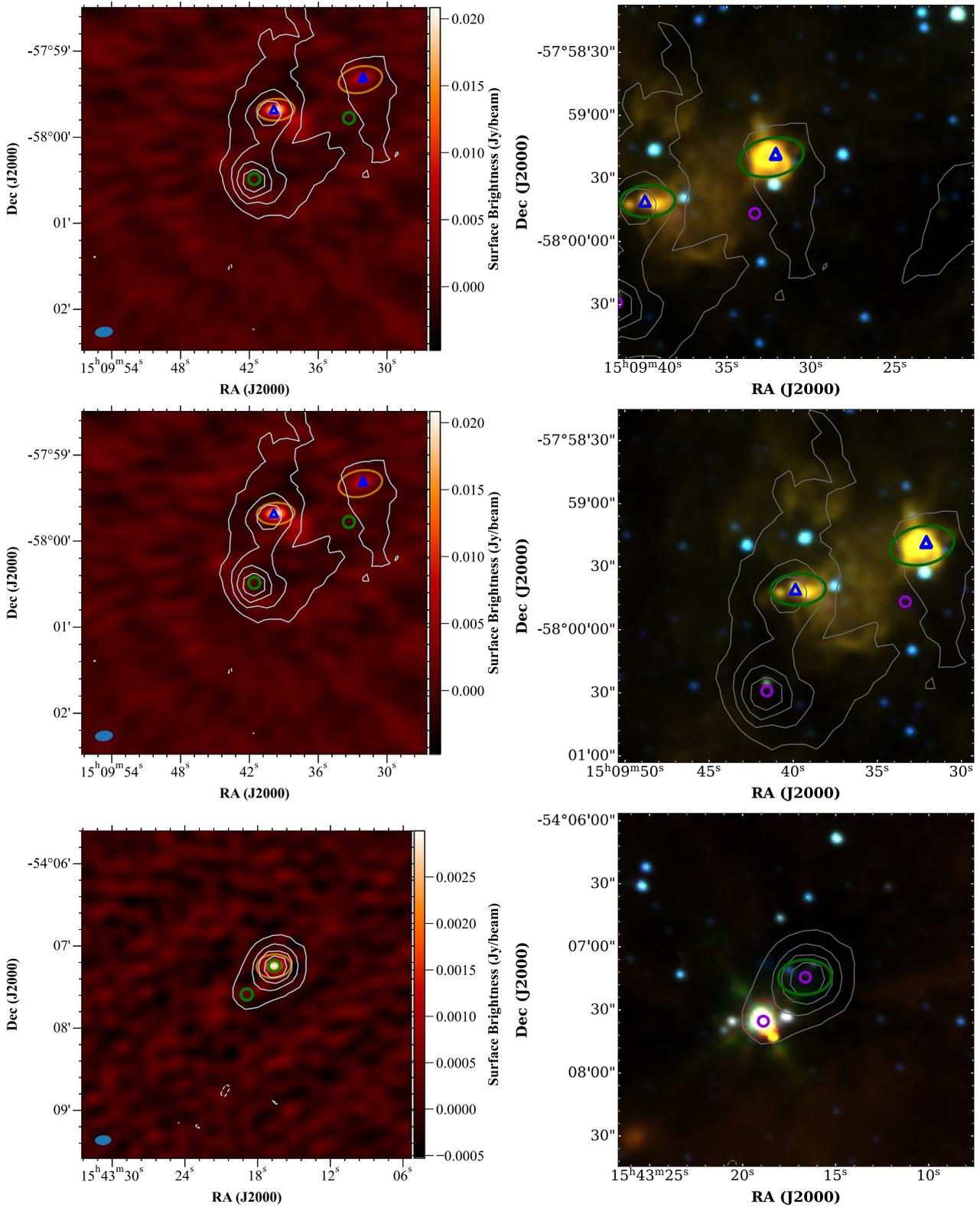

**Figure A1.** Cont. *Top row*: G320.416+0.116, AGAL320.414+00.116, *Middle row*: G320.427+0.103, AGAL320.427+00.102, *Bottom row*: G326.475+0.703, AGAL326.474+00.702.

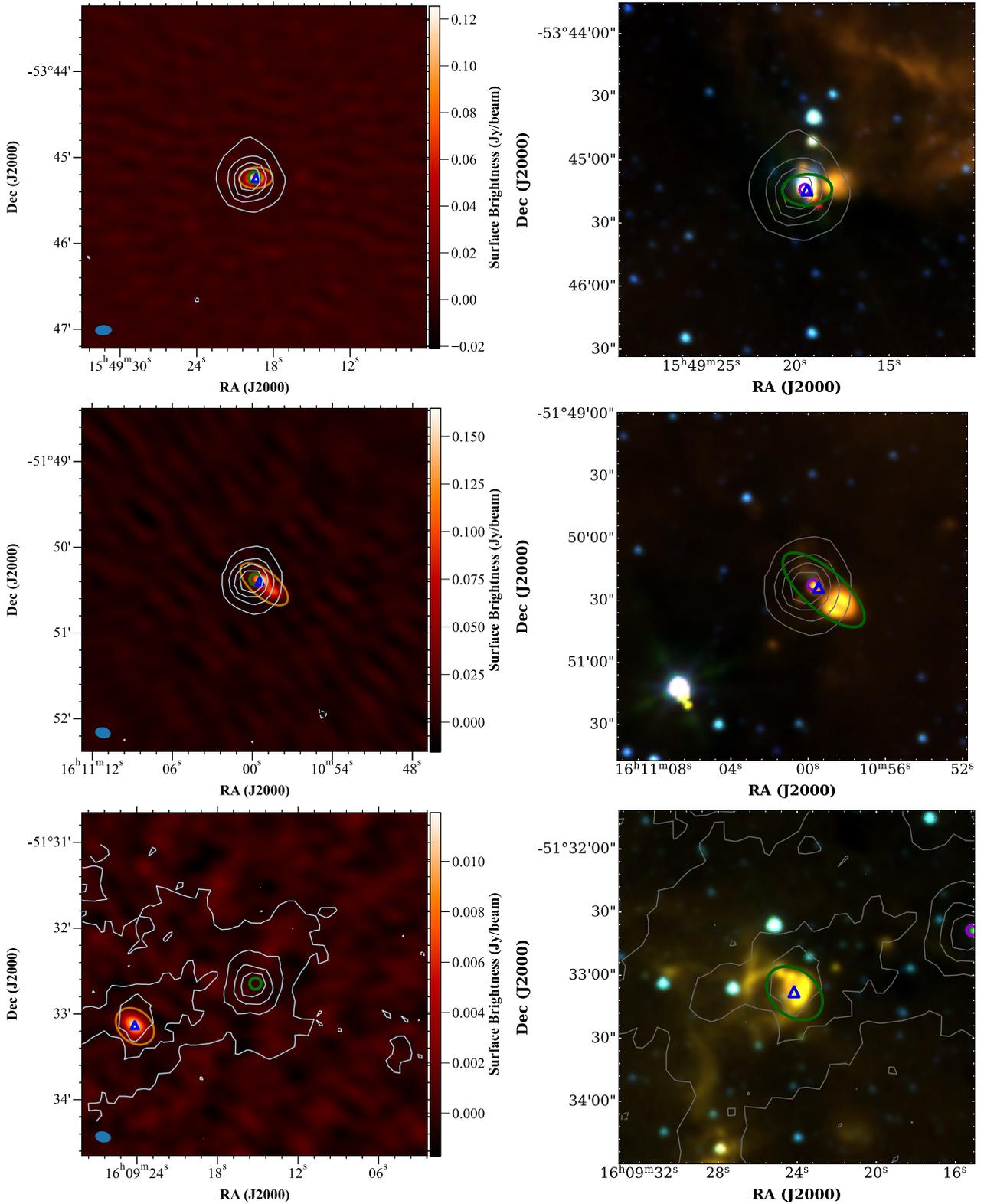

**Figure A1.** Cont. *Top row*: G327.402+0.445, AGAL327.403+00.444, *Middle row*: G331.130−0.243, AGAL331.133−00.244, *Bottom row*: G331.145+0.135, AGAL331.146+00.136.

32    *A.L Patel et al.*

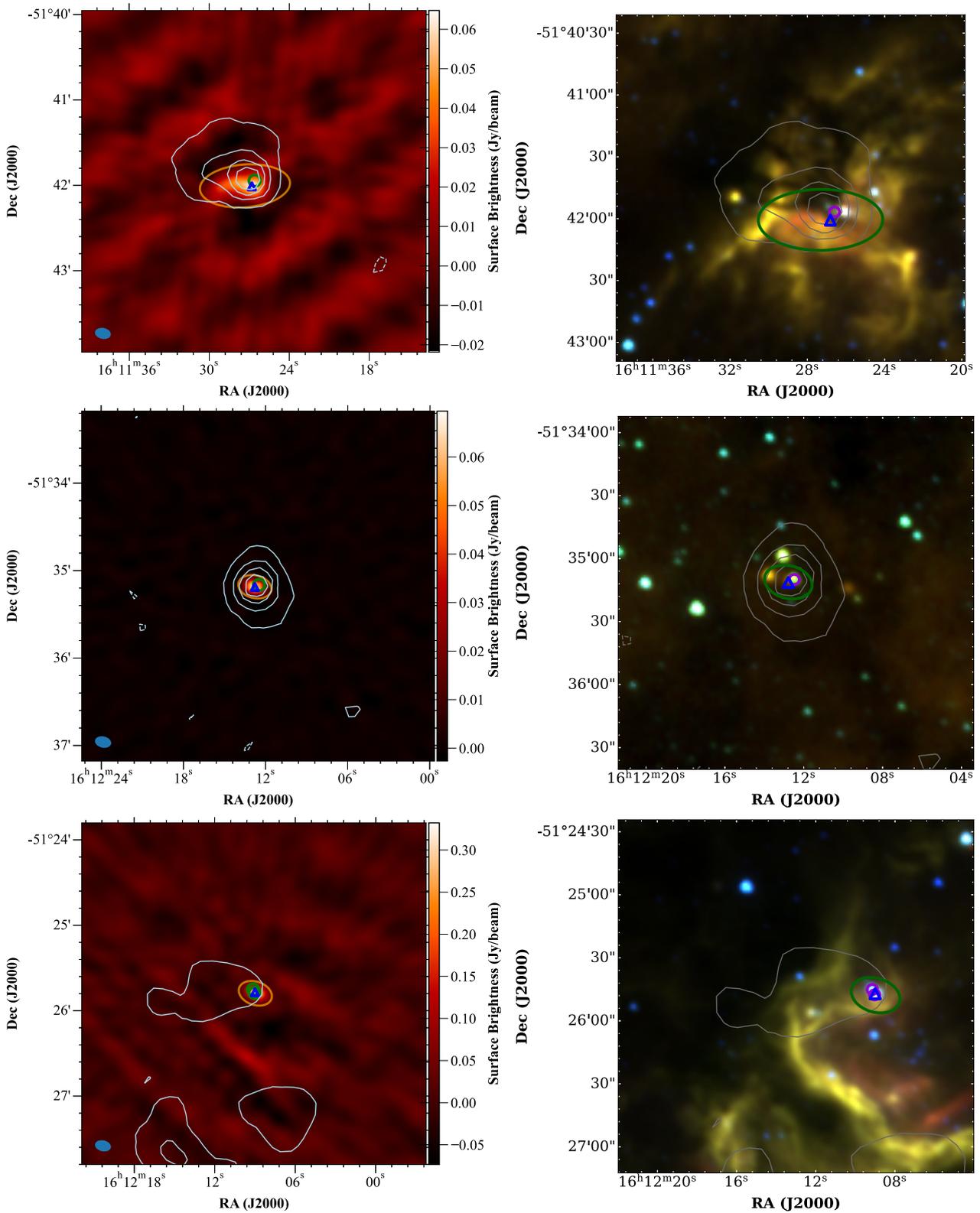

**Figure A1.** Cont. *Top row*: G331.279−0.190, AGAL331.279−00.189, *Middle row*: G331.442−0.187, AGAL331.442−00.187, *Bottom row*: G331.542−0.067, AGAL331.546−00.067.

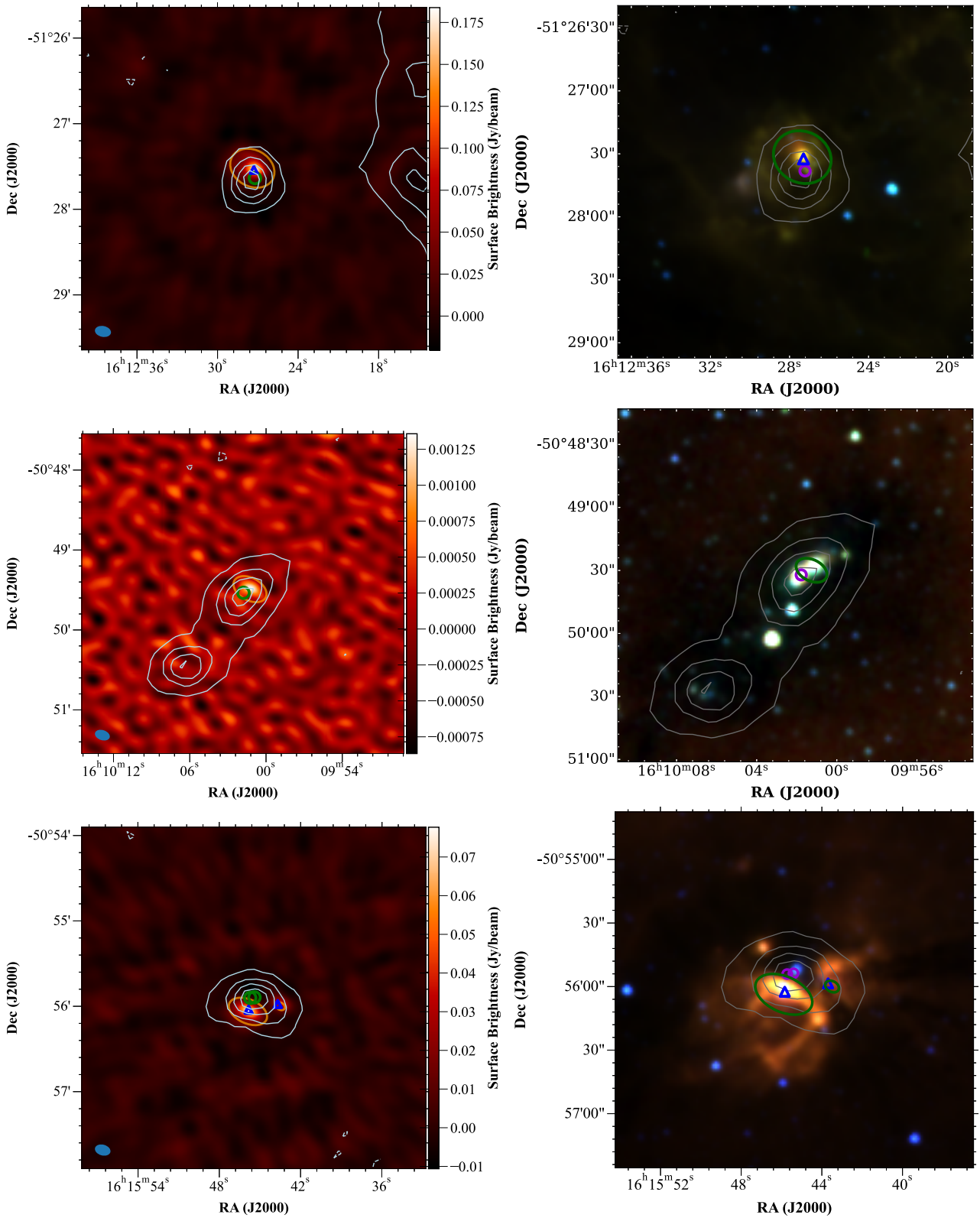

**Figure A1.** Cont. *Top row*: G331.557−0.120, AGAL331.556−00.122, *Middle row*: G331.710+0.604, AGAL331.709+00.602, *Bottom row*: G332.290−0.091, AGAL332.296−00.094.

34  *A.L Patel et al.*

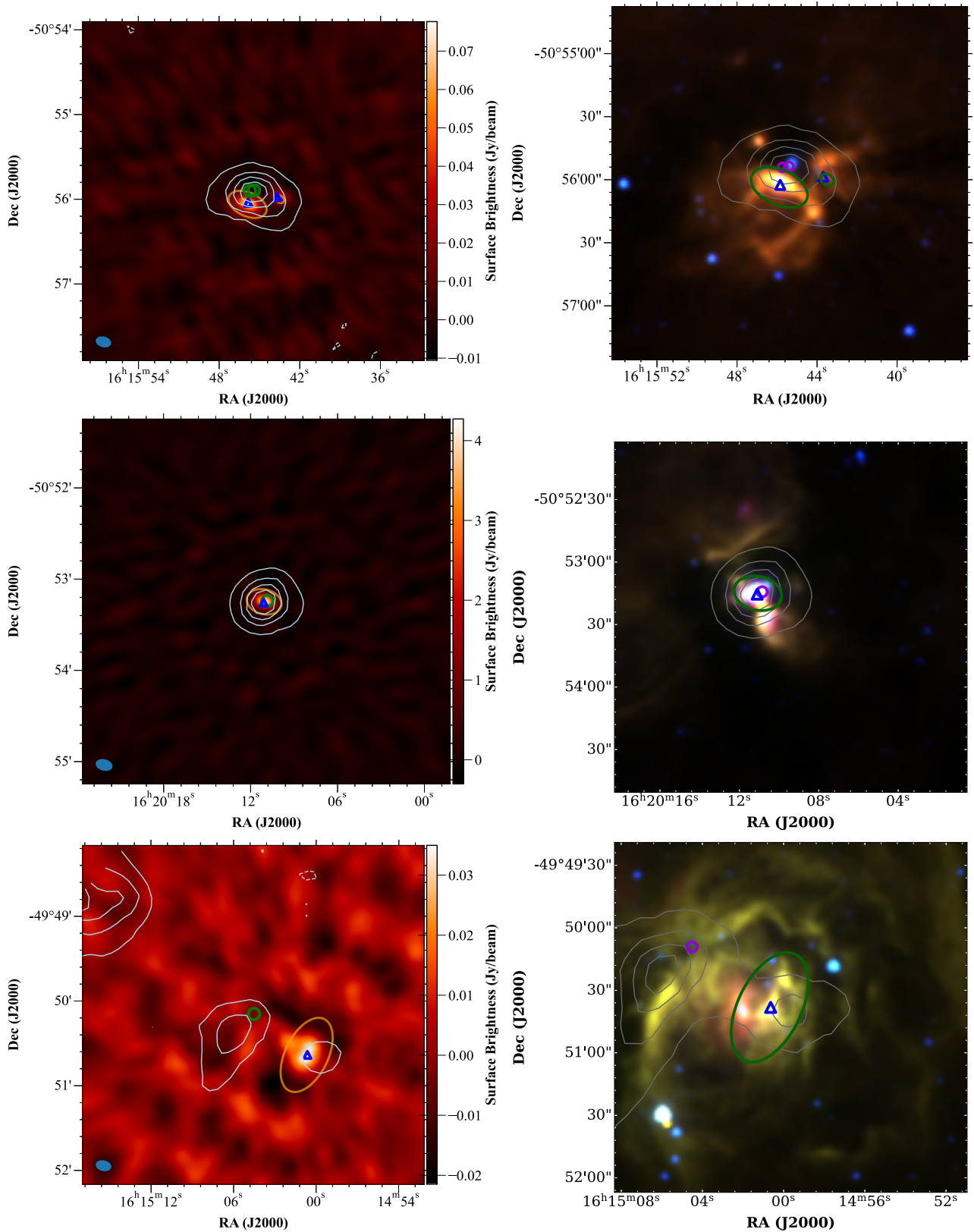

**Figure A1.** Cont. *Top row*: G332.302−0.079, AGAL332.296−00.094, *Middle row*: G332.826−0.549, AGAL332.826−00.549, *Bottom row*: G332.962+0.774, AGAL332.959+00.776.

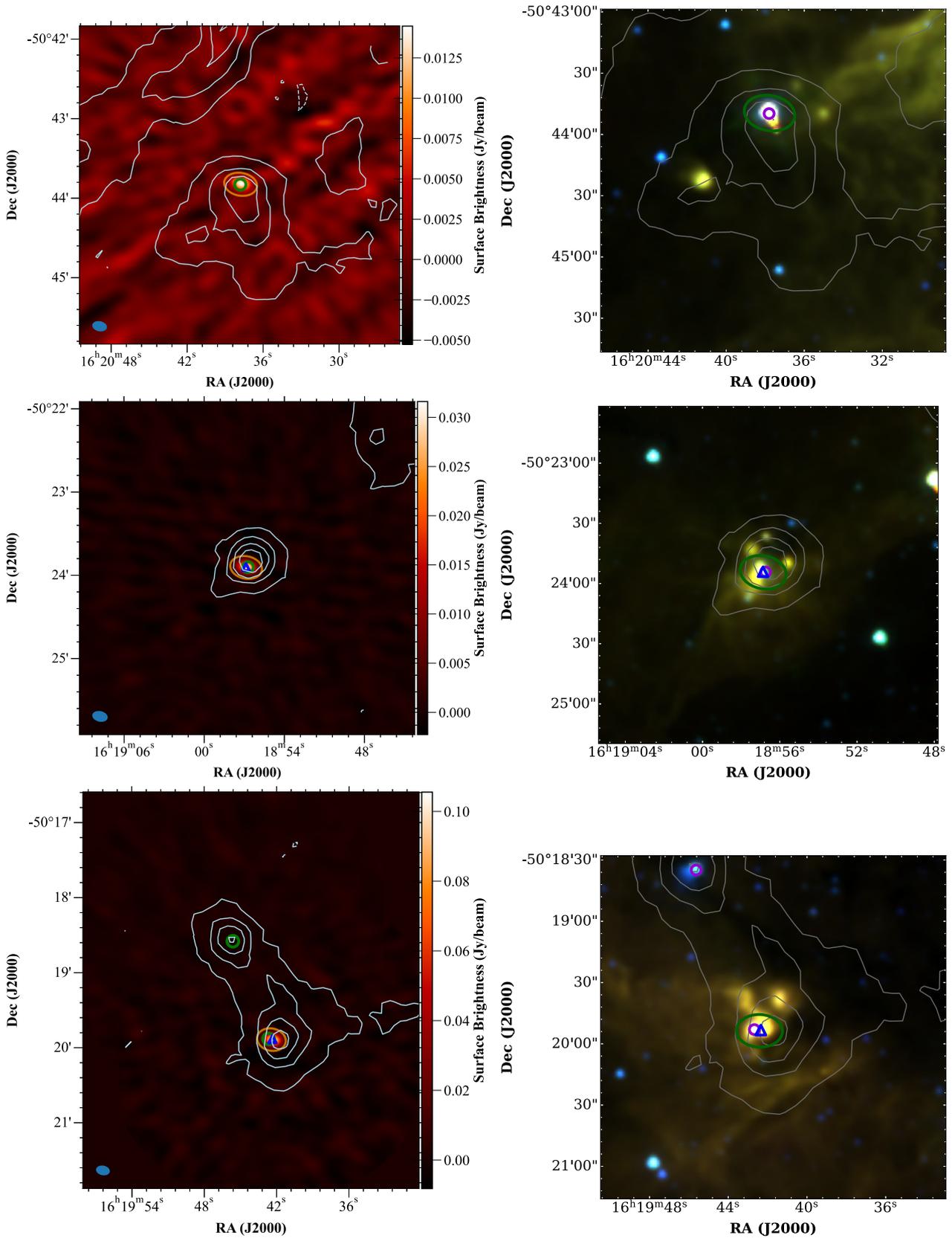

**Figure A1.** Cont. *Top row*: G332.987−0.487, AGAL332.986−00.489, *Middle row*: G333.029−0.063, AGAL333.029−00.061, *Bottom row*: G333.163−0.100, AGAL333.161−00.099.

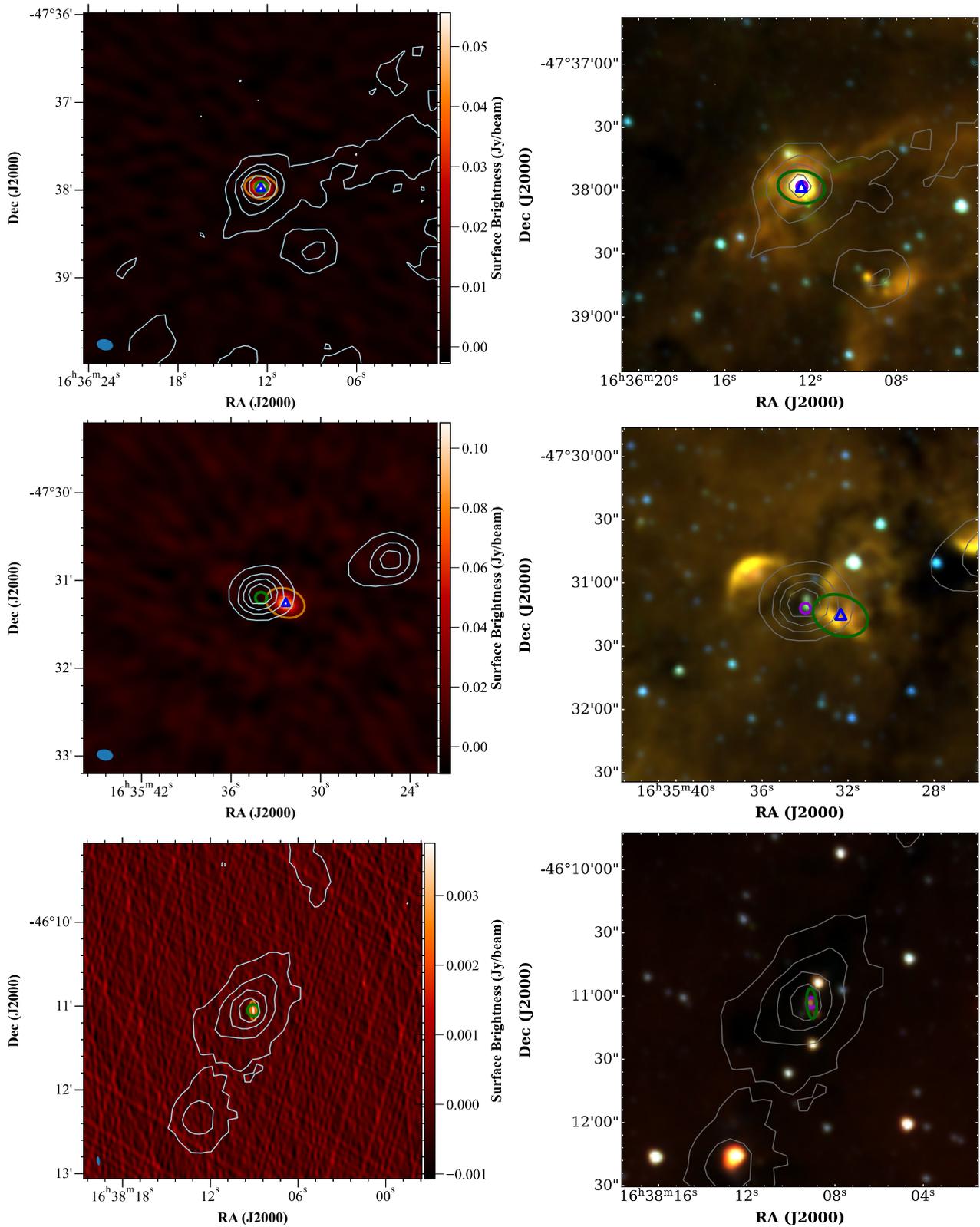

**Figure A1.** Cont. *Top row*: G336.983−0.183, AGAL336.984−00.184, *Middle row*: G336.990−0.025, AGAL336.994−00.027, *Bottom row*: G338.280+0.542, AGAL338.281+00.542.

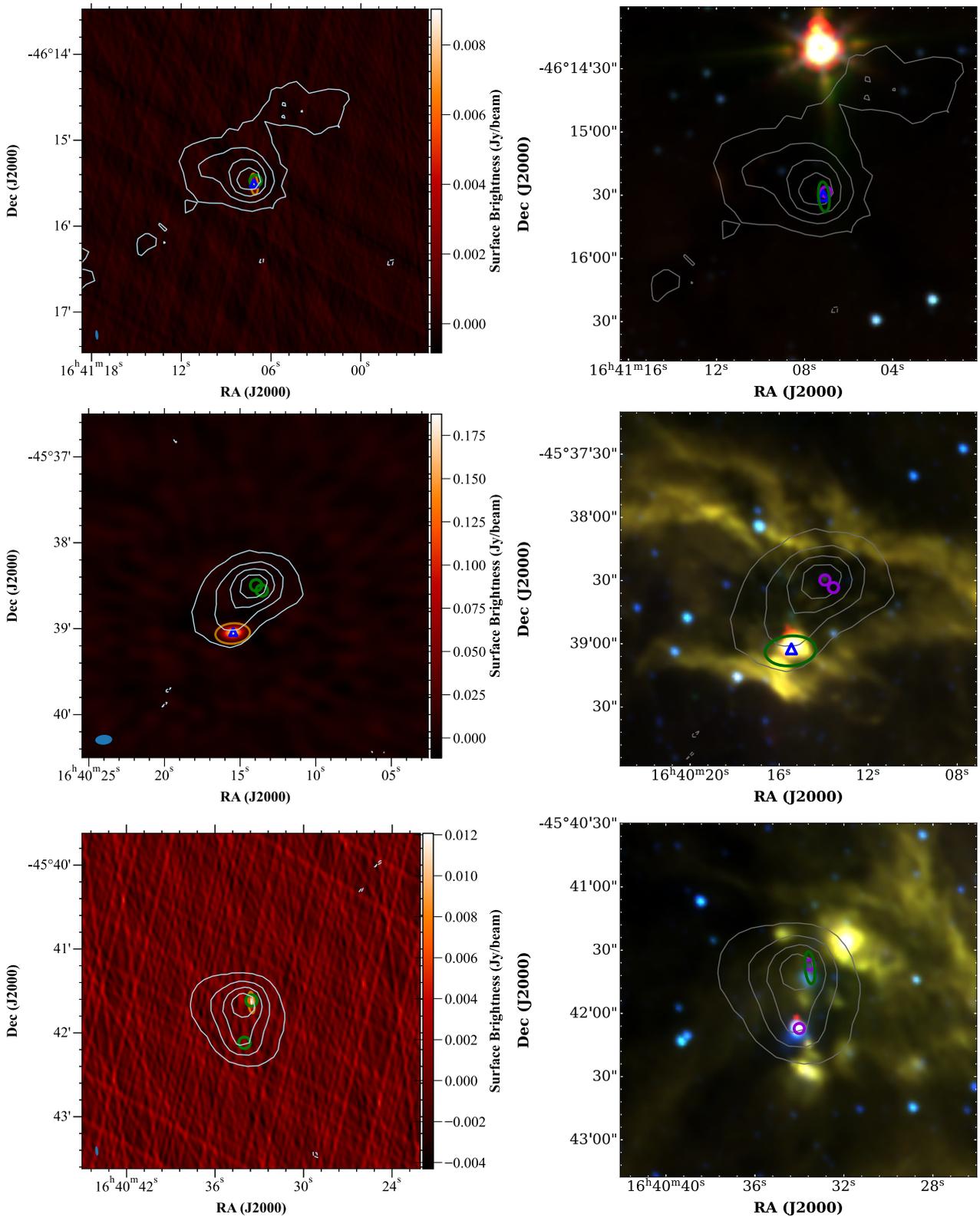

**Figure A1.** Cont. *Top row*: G338.566+0.110, AGAL338.567+00.109, *Middle row*: G338.922+0.624, AGAL338.926+00.634, *Bottom row*: G338.925+0.556, AGAL338.926+00.634.

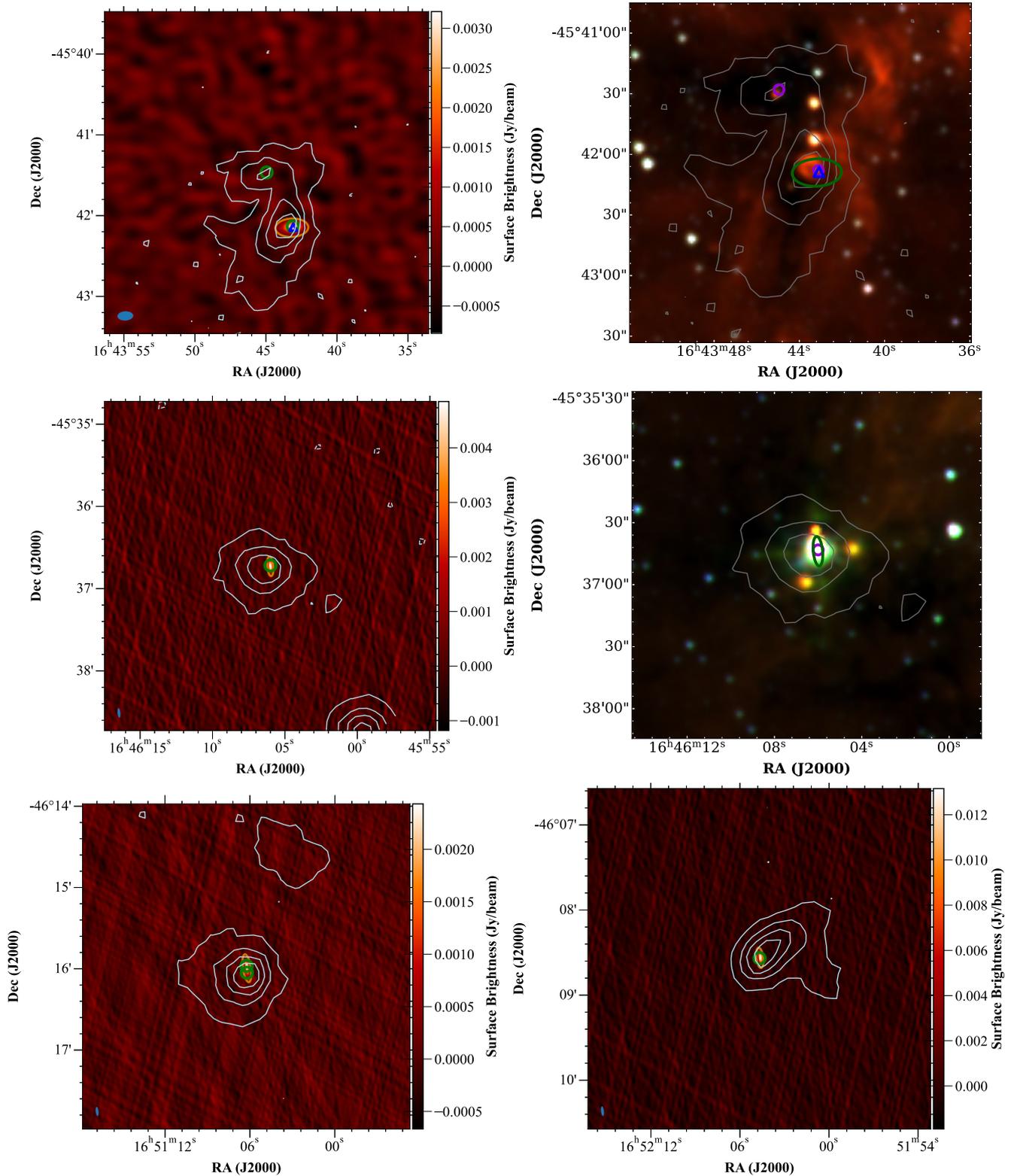

**Figure A1.** Cont. *Top row*: G339.282+0.136, AGAL339.283+00.134, *Middle row*: G339.622−0.121, AGAL339.623−00.122, *Bottom row*: G339.681−1.207, G339.884−1.259.

The positions of the bottom two radio sources are not covered by the WISE−IRAC wavebands.

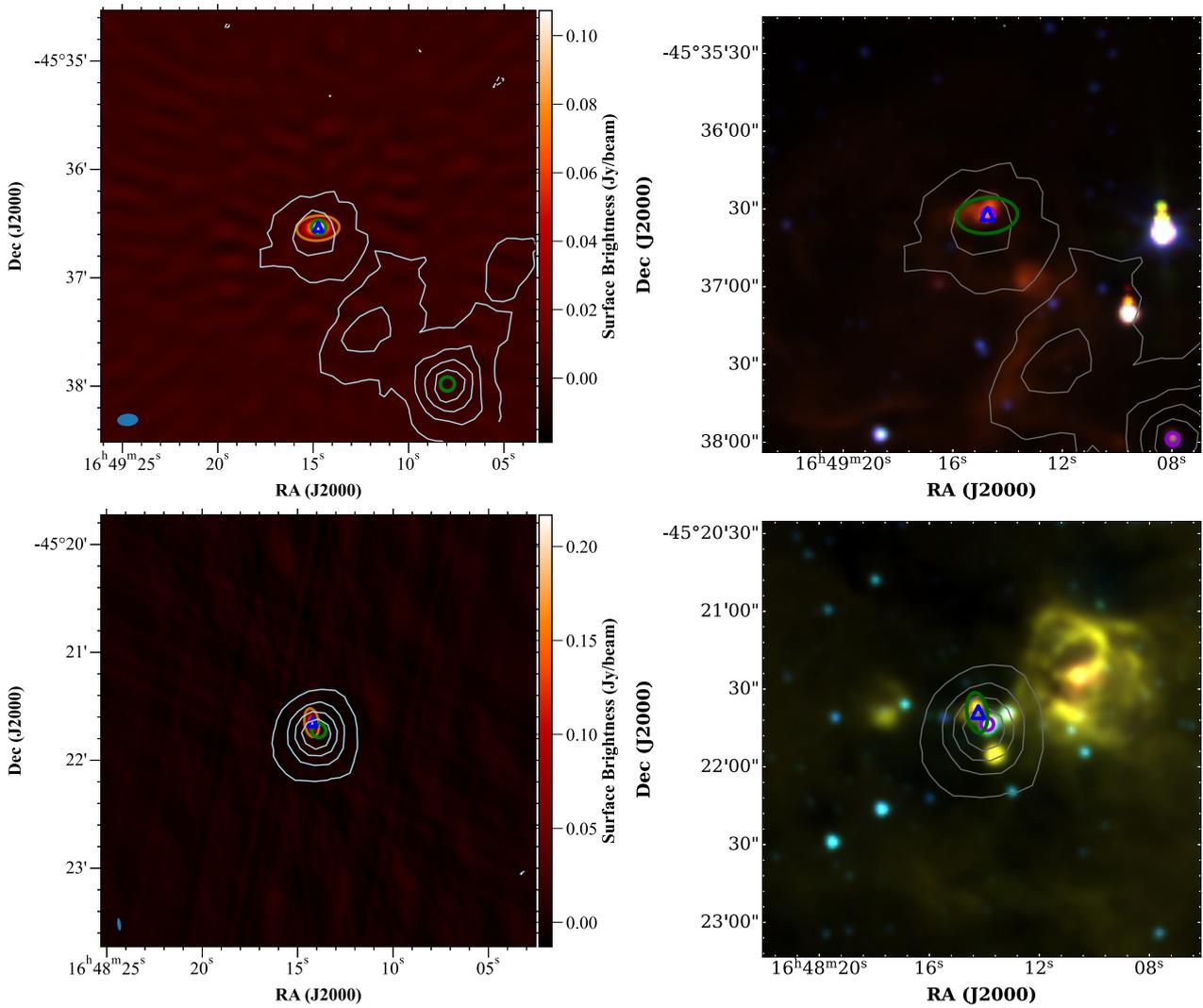

**Figure A1.** Cont. *Top row*: G339.980−0.539, AGAL339.979−00.539, *Bottom row*: G340.056−0.244, AGAL340.054−00.244.

40  *A.L Patel et al.*

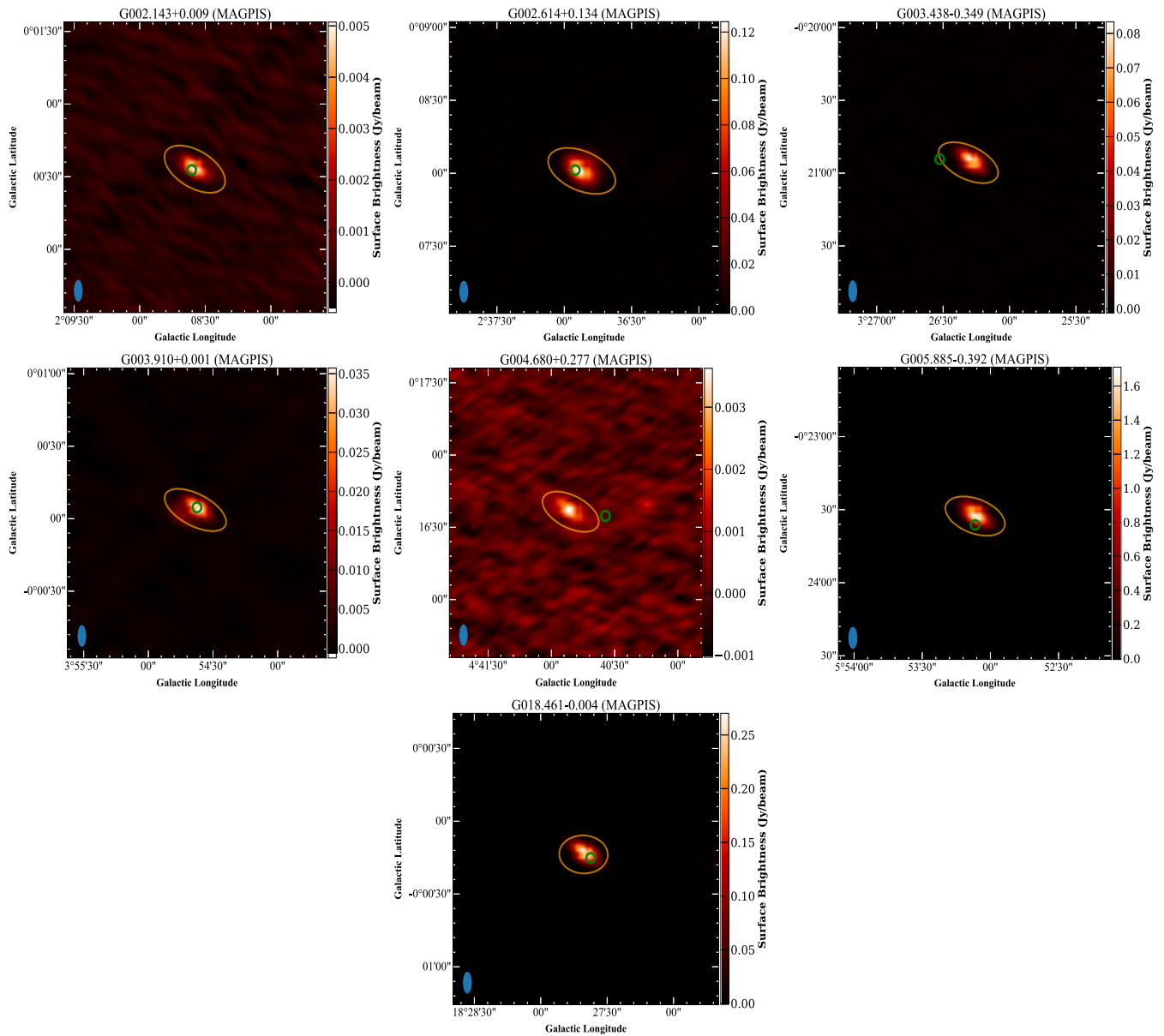

**Figure B1.** Radio maps of the 5 GHz MAGPIS detections. The orange ellipse represents the size and position of our 23.7 GHz data. The green circle indicates the position of any MMB methanol maser(s). The filled blue ellipse in the bottom left hand corner of each image indicates the size of the synthesised beam. The radio name and survey is given on the top of each figure

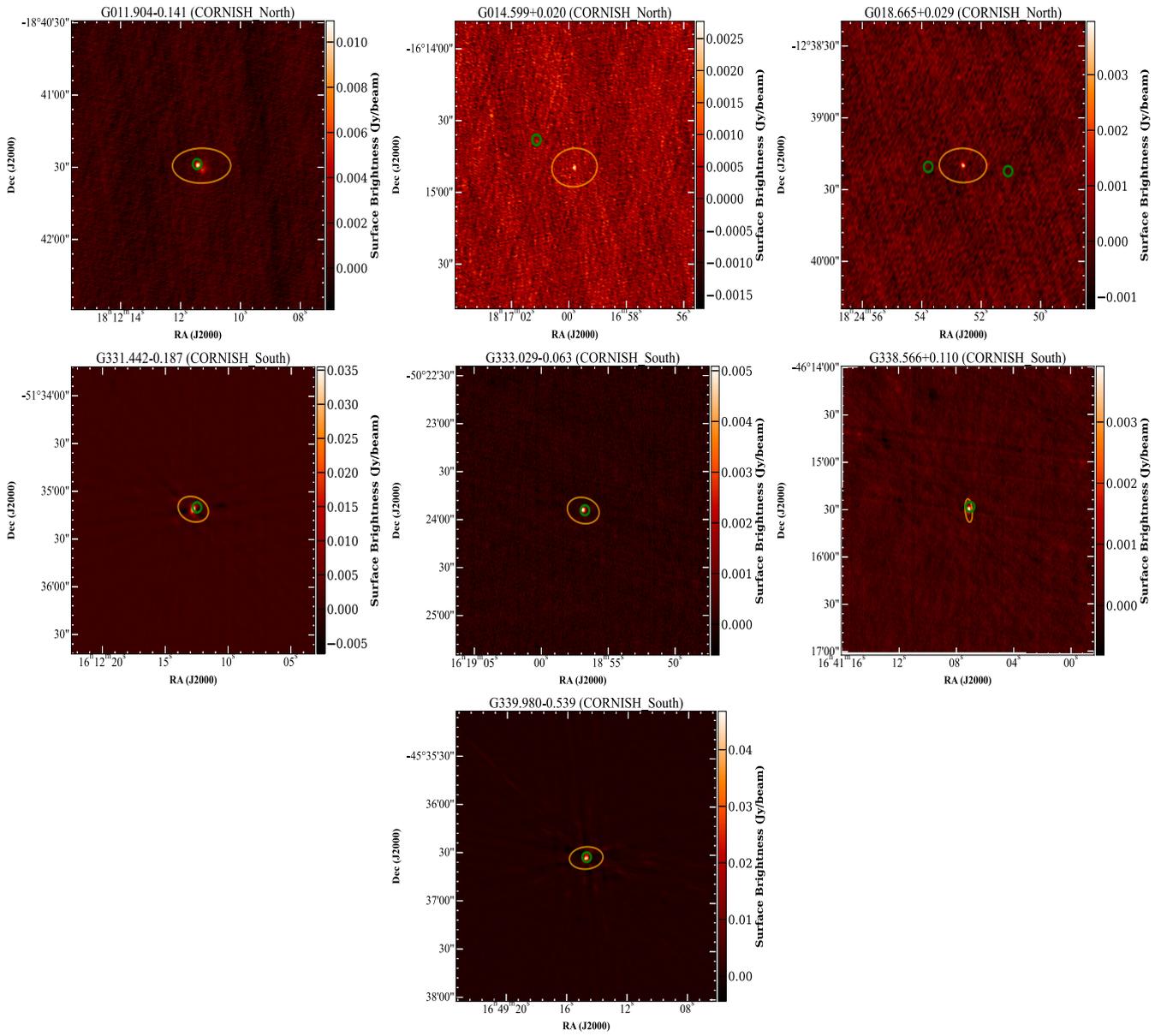

**Figure B2.** Radio maps of the 5 GHz CORNISH detections. The orange ellipse represents the size and position of our 23.7 GHz data. The green circle indicates the position of any MMB methanol maser(s). The radio name and survey is given on the top of each figure

42  *A.L Patel et al.*

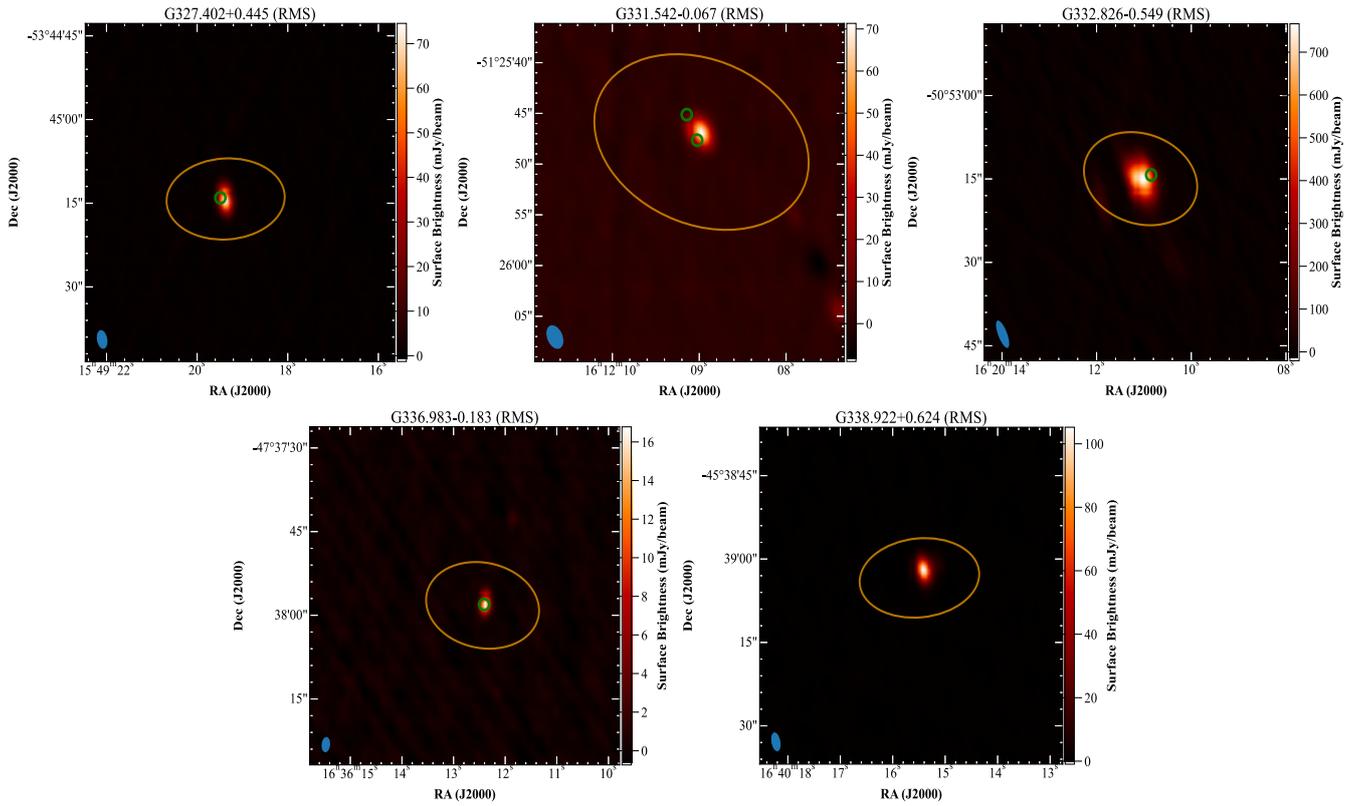

**Figure B3.** Radio maps of the 5 GHz RMS detections. The orange ellipse represents the size and position of our 23.7 GHz data. The green circle indicates the position of any MMB methanol maser(s). The filled blue ellipse in the bottom left hand corner of each image indicates the size and orientation of the synthesised beam. The radio name and survey is given on the top of each figure

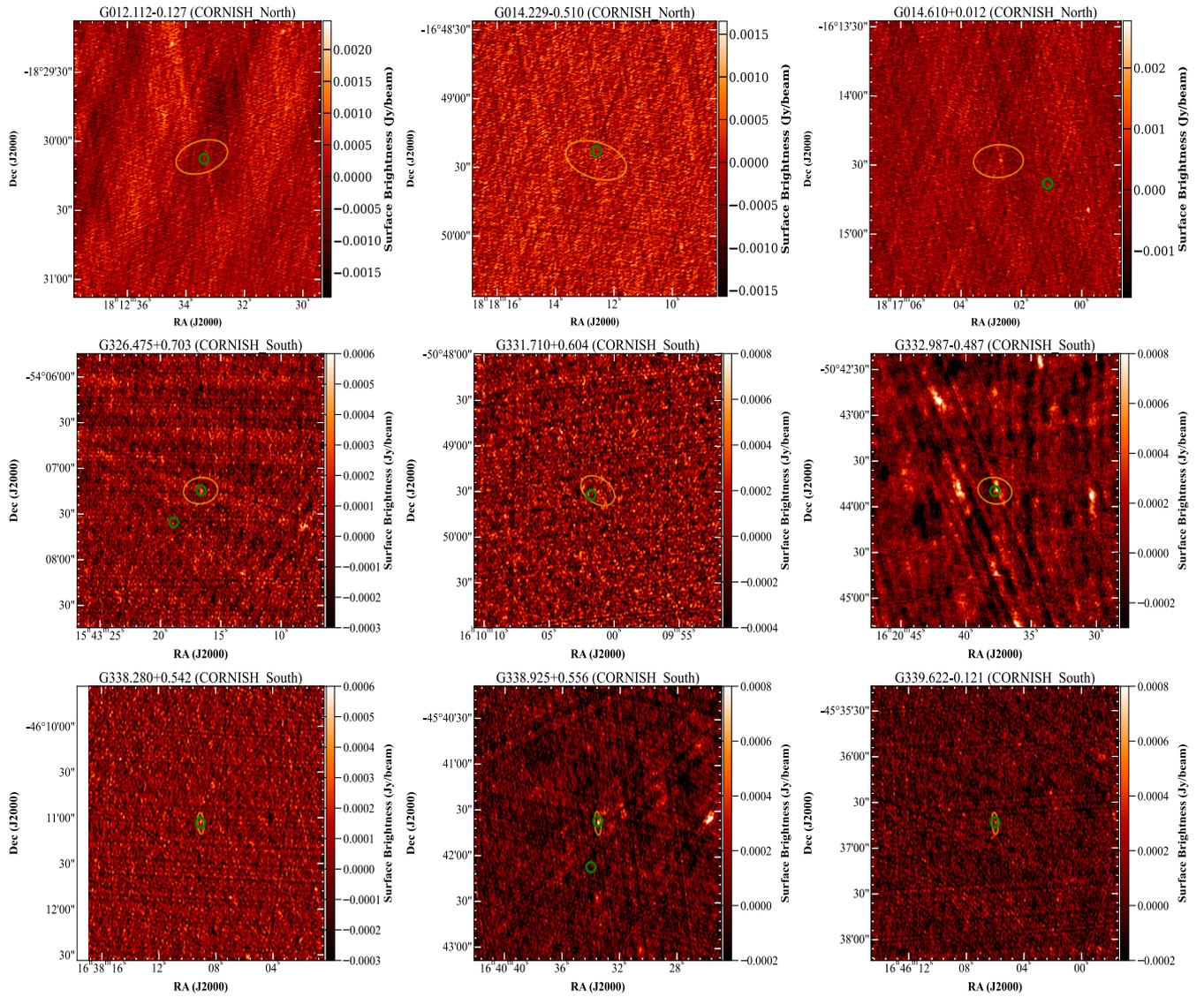

**Figure B4.** Radio maps of the nine, 5 GHz non-detections. The orange ellipse represents the size and position of our 23.7 GHz data. The green circle indicates the position of any MMB methanol maser(s). The radio name and survey is given on the top of each figure.